\documentclass[useAMS,usenatbib]{mn2e}
 
\usepackage{graphics,float}                
\usepackage{epsf}
 
\usepackage{amsmath}                                
\usepackage{amssymb}
\usepackage{picinpar}                               
\usepackage{relsize,setspace,subfigure}
\usepackage{tabularx}
\usepackage[innercaption]{sidecap}                  
\usepackage[figuresright]{rotating}
\usepackage{url}
 
\usepackage{natbib}
\usepackage{aas_macros}

\def\etal{{\it et al}.}

\def\d3k{{\displaystyle {\rm d}{\bf k} \over \displaystyle (2\pi)^3}}
\def\Mpch{~h^{-1} {\rm Mpc}}

\title[Multiscale Phenomenology of the Cosmic Web]{Multiscale Phenomenology of the Cosmic Web}
 
\author[Arag\'on-Calvo, van de Weygaert \& Jones]{Miguel A. Arag\'on-Calvo$^{1,2}$
,Rien van de Weygaert$^2$, Bernard J.T. Jones$^2$\\
  $^{1}$The Johns Hopkins University, 3701 San Martin Drive, Baltimore, MD 21218, USA\\
  $^{2}$Kapteyn Astronomical Institute, University of Groningen, P.O.
  Box 800, 9700 AV, Groningen, The Netherlands.}
 
\begin{document}
 
\date{Accepted .... Received ...; in original form ...}
 
\pagerange{\pageref{firstpage}--\pageref{lastpage}} \pubyear{2008}
 
\maketitle
 
\label{firstpage}
 
\begin{abstract}
We analyze the structure and connectivity of the distinct morphologies that 
define the Cosmic Web. With the help of our Multiscale Morphology Filter (MMF), 
we dissect the matter distribution of a cosmological $\Lambda$CDM N-body computer 
simulation into cluster, filaments and walls. The MMF is ideally suited to 
adress both the anisotropic morphological character of filaments and sheets, 
as well as the multiscale nature of the hierarchically evolved cosmic matter distribution. 
The results of our study may be summarized as follows: 
\textbf{i)}.- While all morphologies occupy a roughly well defined range in density,
this alone is not sufficient to differentiate between them given their overlap. 
Environment defined only in terms of density fails to incorporate the intrinsic dynamics 
of each morphology. This plays an important role in both linear and non linear interactions between haloes. 
\textbf{ii)}.- Most of the mass in the Universe is concentrated in filaments, narrowly followed 
by clusters. In terms of volume, clusters only represent a minute fraction, and filaments 
not more than $9\%$. Walls are relatively inconspicous in terms of mass and volume. 
\textbf{iii)}.- On average, massive clusters are connected to more filaments than low mass 
clusters. Clusters with $M \sim 10^{14}$ M$_{\odot}$ h$^{-1}$ have on average two connecting
filaments, while clusters with $M \geq 10^{15}$ M$_{\odot}$ h$^{-1}$ have on average five connecting
filaments.
\textbf{iv)}.- Density profiles indicate that the typical width of filaments is 2$\Mpch$. Walls
have less well defined boundaries with widths between 5-8 Mpc h$^{-1}$. In their 
  interior, filaments have a power-law density profile with slope $\gamma \approx -1$, corresponding 
  to an isothermal density profile. 
\end{abstract}
 
 
\section{Introduction}
\label{sec:introduction}
The Megaparsec matter distribution in the Universe represents a dynamical system of great structural and 
topological complexity, the Cosmic Web.

Early attempts to map the large scale distribution of galaxies in the universe \citep{Gregory78,Geller89,
Lapparent86,Shectman96} revealed that galaxies are far from being evenly distributed across the nearby 
Universe. On the contrary, the mass distribution delineated by galaxies seems to form an
intricate network of compact and dense associations interconnected by tenuous ``bridges" or ``filaments" surrounded 
by surprisingly vast empty regions \citep{Kirshner81}. These preliminary studies suggested that 
the universe on the large scales could be described as a cellular system \citep{Joeveer78} or a 
\textit{Cosmic Web} \citep{Bondweb96}. This has been confirmed in recent times by large galaxy surveys 
like the 2dFGRS \citep{Colless03}, the Sloan Digital Sky Survey \citep[e.g.]{Tegmark04} and the 2MASS redshift 
survey \citep{Huchra05}. 

The advent of these large 3-D maps of the Local Universe unveiled a cosmos of considerable richness and 
complexity, featuring intricate filamentary structures. These structures can be seen on scales from a few 
megaparsecs up to tens and even hundreds of megaparsecs. They include immense elongated and semi-planar 
patterns \citep{Diaferio97,Tittley01,Stevens04,Ebeling04,Pimbblet04,Pimbblet04b,Bharadwaj04,Pimbblet05b} 
and includes huge wall-like structures like the Coma Great Wall \citep{Geller89} and the Sloan Great 
Wall \citep{Gott05}, with its size of more than 400$\Mpch$ the largest known structure in the 
nearby Universe. Similar weblike structures have also been discovered at high redshifts 
\citep{Broadhurst90,Ouchi04,Cohen96}. 

\subsection{Gravitational Formation}
The Cosmic Web can be seen as the most salient manifestation of the anisotropic nature of gravitational 
collapse, the motor behind the formation of structure in the cosmos \citep{Peebles80}. N-body computer 
simulations have profusely illustrated how a primordial field of tiny Gaussian density perturbations transforms 
into  a pronounced and intricate filigree of filamentary features, dented by dense compact clumps at the nodes of 
the network \citep[][]{Jenkins98,Colberg05,Springel05b,Dolag06}. The filaments connect into the 
cluster nodes and act as the transport channels along which matter flows into the clusters. 

Fundamental understanding of anisotropic collapse on cosmological scales came with the seminal study 
by \citet{Zeldovich70}, who recognized the key role of the large scale tidal force field in shaping the 
Cosmic Web \citep[also see][]{Icke73}. In addition to the anisotropic nature of gravitational collapse, 
the multiscale character of the cosmic mass distribution is also an important characteristic signature of 
the gravitational formation of structure \citep{Zeldovich70,Icke73,Eisenstein95,Bondweb96,Eisenstein97,
WeyBond08a,Shandarin09}. Cosmic structure formation is a hierarchical process as a result of the amplitude distribution of 
fluctuations over the different scales. Small scale fluctuations in scenarios with a primordial 
power spectrum $P(k) \propto k^n$, where $n > -3$, have a larger amplitude than the ones 
on larger scales. As a result, collapsed clumps of matter will aggregate into larger systems 
and eventually merge to form even larger structures. 

The description of the Megaparsec matter distribution as an interconnected network or a
cosmic web is not a coincidence. While the Zel'dovich approximation describes the 
evolution of a cellular distribution made up of pancakes, it does not really offer 
an explicit dynamical expalnation for the observed connectivity between morphologies. Early computer 
simulations already indicated the close connection between each morphological component, namely that clusters
sit at the intersection of filaments and filaments are formed at the intersection of walls 
\citep{Doroshkevich80,Melott83,Pauls95,Shapiro83,Sathyaprakash96}. \citet{Bondweb96} introduced 
the ``Cosmic Web'' theory, which provides a natural explanation for the emergence of the filamentary 
network as well as the relation between the morphological components of the Cosmic Web. 
The theory emphasizes the close relation between the peaks in the density field and the overall 
weblike network: \textit{knowledge of the tidal field at a few relevant locations in a region 
provides all the information needed to predict the resulting large scale matter configuration}. 
In the primordial density field this can be traced back to the simple quadrupolar 
pattern in the density distribution implied by a local shear configuration \citep[see][]{WeyEdb96,WeyBond08a}. 
This distribution will naturally evolve into a cluster-filament-cluster configuration, the structural basis of the 
Cosmic Web.

\begin{figure}
  \centering
  \includegraphics[width=0.45\textwidth]{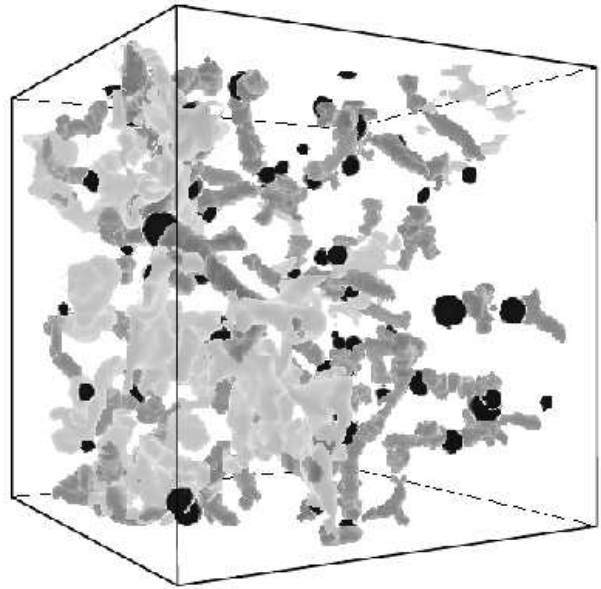}
    \caption{MMF segmentation of the mass distribution in the Cosmic Web. The cosmic web is delineated by filaments 
             (dark gray) and walls (light gray). Clusters (dark grey) are located at the intersection of filaments. 
             Each of these elements is indicated by isodensity contours (on a Gaussian scale of $R_f=2\Mpch$.  
             Only the largest structures are shown for clarity.}
  \label{fig:cosmic_web_surfaces}
\end{figure} 

\subsection{Web Analysis}
Despite the multitude of elaborate qualitative descriptions it has remained a major challenge to characterize the 
structure, geometry and topology of the Cosmic Web. Quantities as basic and general as the mass and volume content 
of clusters, filaments, walls and voids are still not well established or defined. Since there is not yet a common 
framework to objectively define filaments and walls, the comparison of results of different studies concerning properties 
of the filamentary network -- such as their internal structure and dynamics, evolution in time, and connectivity 
properties -- is usually rendered cumbersome and/or difficult. 

The overwhelming complexity of the individual structures as well as their connectivity, the lack of structural symmetries, its 
intrinsic multiscale nature and the wide range of densities that one finds in the cosmic matter distribution has prevented the 
use of a simple and straightforward tool box. Over the years, a variety of heuristic measures were proposed to analyze specific 
aspects of the spatial patterns in the large scale Universe. Only in recent years these have lead to a more solid and well-defined 
machinery for the description and quantitative analysis of the intricate and complex spatial patterns of the Cosmic Web. Nearly without 
exception, these methods borrow extensively from other branches of science such as image processing, mathematical morphology, computational 
geometry and medical imaging. 

The connectedness of elongated supercluster structures in the cosmic matter distribution was first probed by means 
of percolation analysis, introduced and emphasized by Zel'dovich and coworkers \citep{Zeldovich82,Shandzeld89,Shandarin04,
Shandarin09b}, while a related graph-theoretical construct, the minimum spanning tree of the galaxy distribution, was 
extensively probed and analysed by Bhavsar and collaborators \citep{Barrow85,Graham95,Colberg07} in an attempt to develop 
an objective measure of filamentarity. Finding filaments joining neighbouring clusters has been tackled, using quite different 
techniques, by \citet{Colberg05} and by \citet{Pimbblet05}. 

More general filament finders have been put forward by a number of authors. Following specific physical criteria, 
\cite{Gonzalez09} recently proposed an interesting and promising combination of a tessellation-based density estimator 
and a dynamical binding energy criterion. A thorough mathematical nonparametric formalism involving the medial axis of 
a point cloud, as yet for 2-D point distributions, has recently been proposed by \cite{Genovese10}. It is based on a 
geometric representation of filaments as the medial axis of the data distribution. Also solidly rooted within a geometric 
and mathematical context is the more generic geometric inference formalism developed by \cite{Chazal09}. It allows the 
recovery of geometric and topological features of the supposedly underlying density field from a sampled point cloud 
on the basis of distance functions. \citet{Stoica05,Stoica07,Stoica10} use a generalization of the classical Candy model to 
locate and catalogue filaments in galaxy surveys. This approach has the advantage that it works directly with the original 
point process and does not require the creation of a continuous density field. However, computationally it is very demanding. 

The more recent formalisms that are intent on characterizing the full range of weblike formalisms usually exploit the 
morphological information in the gradient and Hessian of the density field or potential field, i.e. the 
tidal field \citep[see e.g.][]{Sousbie08a,Aragon07a,Aragon07b,Hahn07a,Hahn07b,Forero08,Bond09,Bond10}. Morse theory 
\citep[see][]{Colombi2000} forms the basis of the {\it skeleton analysis} by \cite{Novikov06} (2-D) and \cite{Sousbie08a} 
(3-D). It identifies morphological features with the maxima and saddle points in the density field and results in an 
elegant and mathematically rigorous tool for filament identification. However, it is computationally intensive, focusses  
mostly on the filaments and is strongly dependent on the smoothing scale of the density field. A more elaborate 
classification scheme on the basis of the manifolds in the tidal field -- involving all morphological features in the 
cosmic matter distribution -- has been proposed by \cite{Hahn07a} \citep[also see][]{Hahn07b,Forero08}. Its great virtue is 
that it is based on the structure of the tidal field, which links it directly to our theoretical understanding of 
the formation and evolution of the Cosmic Web. 

\begin{table*} 
\centering
\begin{tabular} {l l l l l l l l l}
\hline
\hline
&&&&&&&&\\
Name  & Box size          & $\Omega_{m}$ & $\Omega_{\Lambda}$ & $h$ & $\sigma_8$ & N$_{part}$ & M$_{part}$    & Softening\\
& [$\Mpch$] & & & & & & [M$_{\odot}$] & [kpc]\\
&&&&&&&&\\
\hline
&&&&&&&&\\ 
150$_{\textrm{High}}$  & 150.0  & 0.3        & 0.7       & 0.7 & 1        &  $512^3$  & $2.09\times10^{9}$ & 18/6 \\
150$_{\textrm{Med}}$   & 150.0  & 0.3        & 0.7       & 0.7 & 1        &  $256^3$  & $1.67\times10^{10}$ & 36/12 \\
150$_{\textrm{Low}}$   & 150.0  & 0.3        & 0.7       & 0.7 & 1        &  $128^3$  & $1.34\times10^{11}$ & 72/24 \\
&&&&&&&&\\
\hline
\end{tabular}
\caption{Parameters of the N-body simulation used in this study.}
\label{tab:n_body_simulations}
\end{table*}

Instead of using the tidal field configuration, one may also try to link directly to the morphology of the density 
field itself. Usually, this allows a more detailed view of the intricacies of the multiscale matter distribution, 
although it is usually more sensitive to noise and more indirectly coupled to the underlying dynamics of structure 
formation than the tidal field morphology. A single scale dissection of the large scale density field into its various 
morphological components based on the has been followed by \cite{Bond09}, and applied to N-body simulations and galaxy 
redshift samples \citep[also see][]{Bond10,Choi10}. 

In this study we follow the more elaborate multiscale formalism of the Multiscale Morphology Filter (MMF), introduced by 
\cite{Aragon07b}. The MMF assigns a morphology of the local density field in terms of its multiscale second order variations 
in the local density field. Instead of restricting the analysis to one particular scale, the MMF explicitly adresses the 
multiscale nature of the cosmic density field by evaluating the density field Hessian over a range of spatial scales and 
determining at which scales and locations the various morphological signatures are most prominent. It represents a complete 
and self-consistent framework that allows us to identify and isolate specific morphologies in an objective way.
A somewhat similar multiscale approach is the Metric Space Technique described by \cite{Wu09}, who applied it to a morphological 
analysis of the DR5 of the SDSS. 

A more recent development is that of the Spineweb procedure \citep{Aragon08,Aragon10}, which traces the various 
features of the cosmic web on pure topological grounds by invoking the {\it Watershed Transform}. The watershed transform 
is a key instrument for the segmentation of a density field, and as such is also ideally suited for tracing 
the boundaries between the identified segments. Spineweb identifies the filaments and sheets with the boundaries of 
watershed basins. The latter are the influence areas in and around cosmic voids. 

\subsection{Cosmic Environment and Galaxy Formation}
One of the main reasons for our interest in outlining the filamentary cosmic web concerns the question 
whether and to what extent the weblike environment influences the properties and evolution of galaxies. Most 
studies of environmental influences limit themselves to the density, but various indications argue for 
a more intricate connection.  

In at least one aspect we may immediately suspect a significant relation between the tidally induced 
morphological nature of the cosmic environment and the galaxy. The tidally induced rotation of galaxies implies 
a link between the galaxy formation process and the surrounding external matter distribution. With the cosmic 
web as a direct manifestation of the large large scale tidal field, we would thereofre expect a connection 
with the angular momentum of galaxies or galaxy halos. The theoretical studies of \citet{Sugerman00} 
and \citet{LeePen00} were important in pointing out that this connection should be visible in the orientation of 
galaxy spins with the surrounding large scale structure. 

Equipped with some of the filament and wall detection techniques described above, recent $N$-body simulations 
have been able to find, amongst others, that the filamentary or sheetlike nature of the 
environment has a distinct influence on the shape and spin orientation of dark matter haloes 
\citep{Aragon07a, Hahn07a, Hahn07b, Paz08, Hahn09, HahnPhd09,Zhang09}. In the case 
of haloes located in large scale walls, they seem to agree that both the spin vector and the major axis 
of inertia lie in the plane of the wall. In the case of the alignment of halos with their embedding filaments, 
\cite{Aragon07a} and \cite{Hahn07a} found evidence for a mass and redshift dependence, which has 
been confirmed by the studies of \cite{Paz08} and \cite{Zhang09}. 

While the Multiscale Morphology Filter \cite{Aragon07b} proved succesfull in elucidating halo shape and spin alignment 
characteristics in filaments and sheets in N-body simulations \citep{Aragon07a,Zhang09}, in recent work \citep[][]{Jones10} 
also succeeded in identifying and tracing filaments in the SDSS survey which contained manifestly aligned galaxies. 
In this paper we use the Multiscale Morphology Filter (MMF) to look in more detail at the intrinsic properties of the 
weblike structures themselves.  

\subsection{Intention and Outline}
We address the Cosmic Web in terms of its basic morphologies -- clusters, filaments and walls -- identified 
on the basis of the Multiscale Morphology Filter (MMF). We will mainly focus on filaments given the fact that they are 
the most prominent components of the cosmic web and largely delineate its outline. Walls are far less prominent, 
more tenuous and highly complex. We will therefore 
pay less attention to them. In some cases we will also include clusters and voids in our analysis but always in 
the context of the filament-wall network. Instead of seeking to provide a comprehensive list of properties of the 
morphologies in the cosmic web, this work presents a general view of the cosmic web from the point of view of their 
morphological components and introduces some tools for their characterization. Some of our results confirm previous findings,
while others have not been presented before. 

This study is organized as follows. In section~\ref{sec:nbody} we describe the cosmological simulation on which this work is 
based, including the resampling of the discrete particle set into a regular grid of density values using the adaptive and morphology 
preserving DTFE technique. Section~\ref{sec:mmf} briefly describes the steps followed in the morphological 
characterization of the Cosmic Web by means of the MMF. Section~\ref{sec:webcomp} contains a qualitative presentation of 
the various morphological components of the Cosmic Web, while the corresponding quantitative inventory in terms of mass, 
volume and density is the subject of section~\ref{sec:webmass}. 

In section~\ref{sec:filament} we proceed to describe the filamentary network
and some of its global properties such as mass function, length distribution,
and density profiles. Finally, we summarize our findings in section~\ref{sec:conclusions}. 

\section{N-body simulations and halo catalogues}
\label{sec:nbody}
The work presented here is based on a cosmological N-body simulation containing 
only dark matter particles. The simulation follows the evolution of a set of 
particles ``tracing'' the underlying density field from a given set of initial conditions
until the present time. We adopted the concordance  $\Lambda$CDM cosmological model
$\Omega_m=0.3$, $\Omega_\Lambda=0.7$, $h=0.7$ and $\sigma_8=1.0$.
Its size (150 $\Mpch$) makes it suitable to study large structures comparable to those seen 
in present galaxy surveys. The large number of particles ($512^3$) allows us 
to achieve a mass resolution of $1.34\times 10^{9}$ M$_{\odot}$ h$^{-1}$ per particle.
The mass resolution and simulation box were chosen as a compromise between
a box large enough to contain a significant amount of large structures and at the same time the 
ability to resolve haloes down to a few times $10^{11}$ M$_{\odot}$ (given the computational 
resources available) . The simulation was performed using the public version of 
the parallel Tree-PM code Gadget2 \citep{Springel05}, running on 8 double processor nodes on 
the Linux cluster at the University of Groningen. Initial conditions at redshift $z=50$ with $512^3$ dark matter particles  were 
generated using the transfer function given by \citet{bbks}.

\begin{figure*}
  \centering
  \includegraphics[width=0.9\textwidth,angle=0]{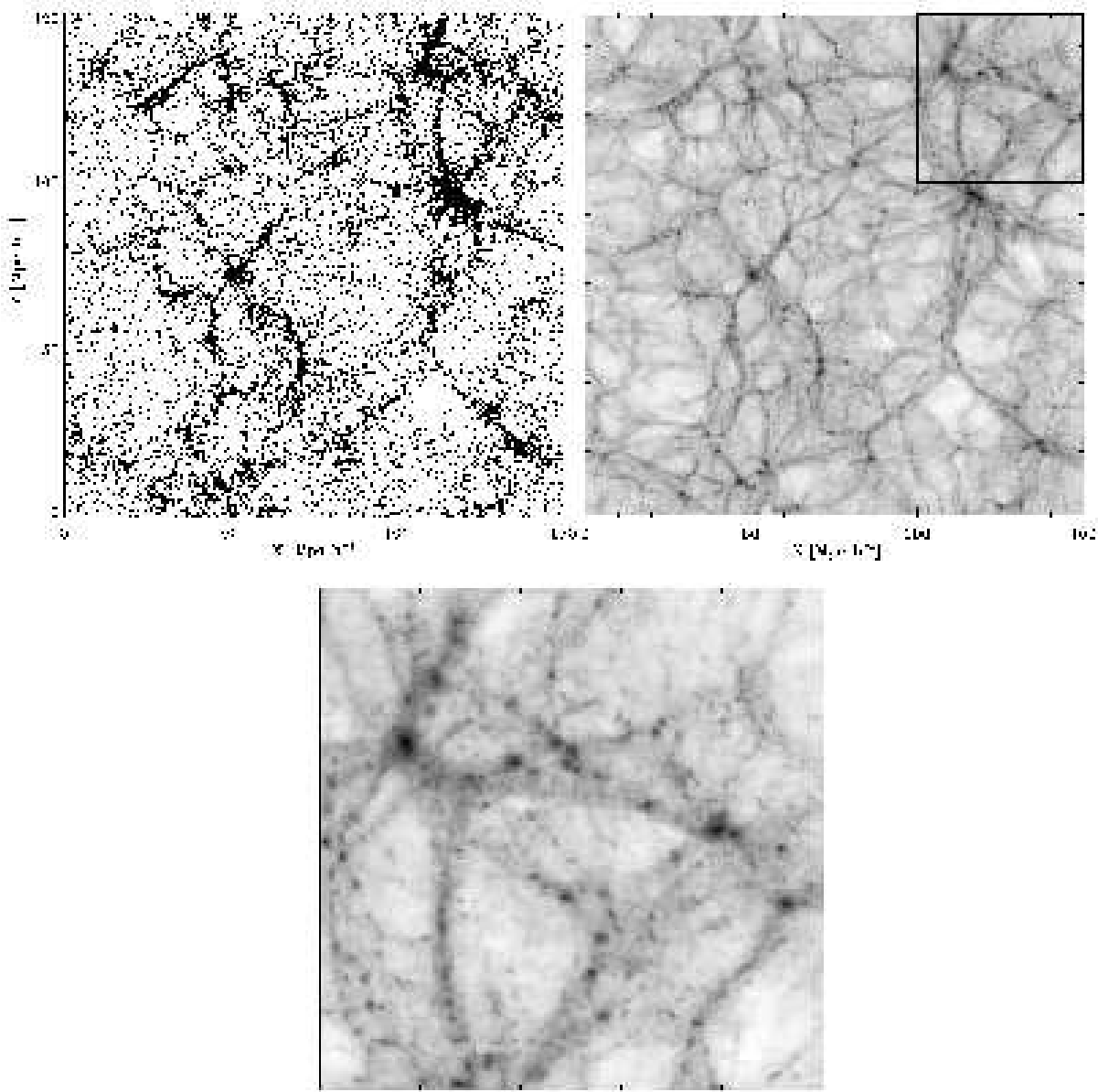}
    \caption{N-body simulation. The particle distribution and density fields of the LCDM simulation 
        which form the data sample for the present study. Particles in a slice of 25 $\Mpch$ along the $z$ axis (top left) and 
        its corresponding DTFE density field (top right) shown in logarithmic scale. The bottom panel shows the zoomed region
        indicated in the top-right corner of the top-right panel. For details of the simulation see text.}
  \label{fig:MMF150H_particles_dtfe}
\end{figure*}
\begin{figure*}
  \centering
  \includegraphics[width=0.9\textwidth,angle=0.0]{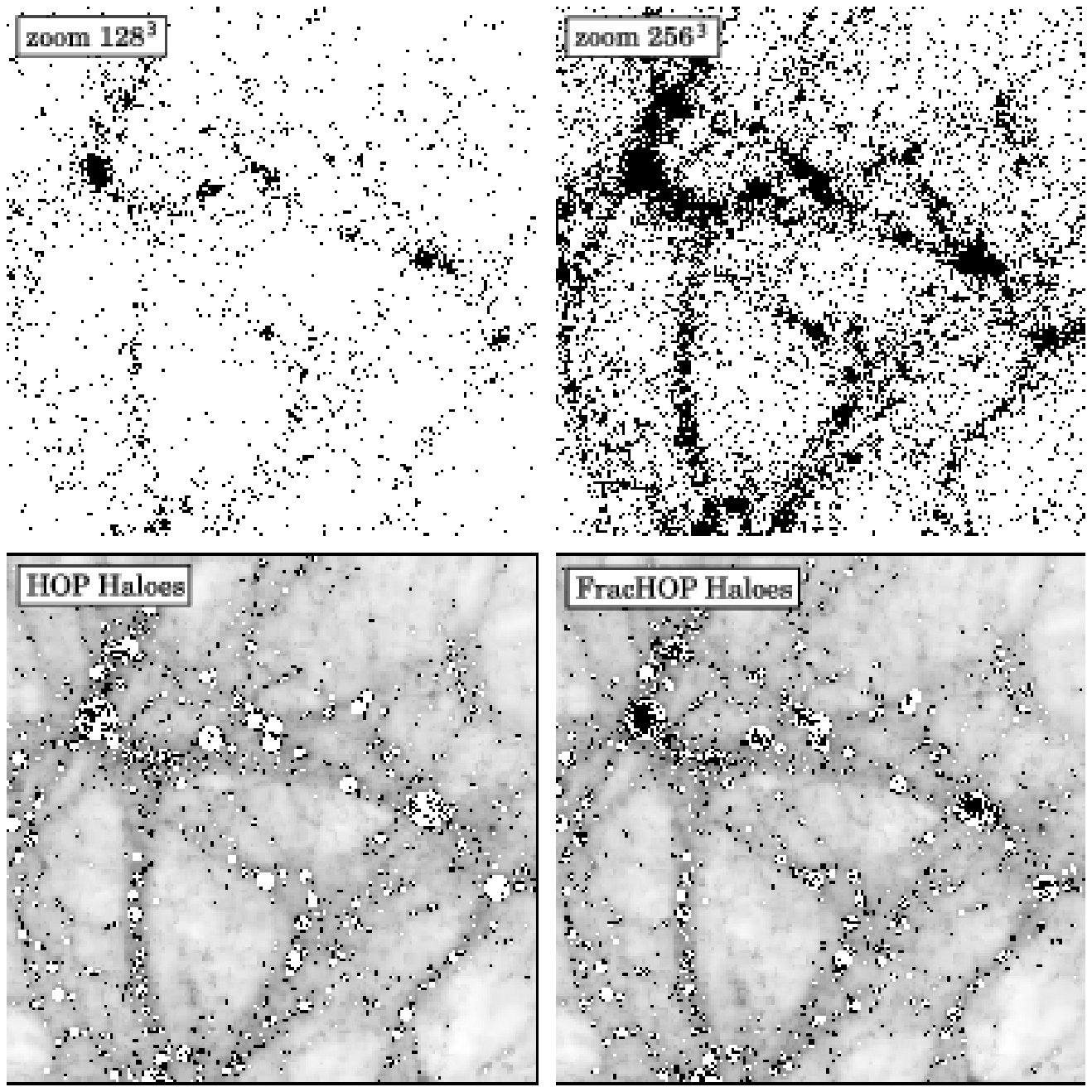}
  \caption{Simulation Particle and Halo distribution. Particles in a slice of 10  $\Mpch$ 
    along the $z$-axis (Top left). The middle panels show particles from the $128^3$ and $256^3$ 
    simulations (left and right respectively)
    in the zoomed region indicated in the top left panel. The lower panels show the Haloes identified
    with HOP (left) and FracHOP (right). The circles are located at the center of mass of the Haloes.
    The radius of the circles are scaled with the mass of the Halo as $r \propto M^{1/3}$.}
  \label{fig:MMF150H_particles_Haloes}  
\end{figure*} 

\subsection{The N-body data}
We stored 20 snapshots starting at redshift 9 in logarithmic intervals of the expansion 
factor until the present time.  Additionally we generated $256^3$ and $128^3$ versions following the 
``averaging" prescription described in \citep{Klypin01}. These lower resolution simulations were used 
to compute the density fields and to get a preliminary impression of the structures present in the 
simulation box (see table 1 and fig.~\ref{fig:MMF150H_particles_Haloes}). 

\begin{itemize}
\item The low resolution version ($n_{part} = 128^3$) is used to compute some properties of filaments such as linear 
density and for visualization purposes. This resolution per particle of this simulation allows
us to resolve the main features of the large scale distribution and at the same time is
sparse enough to allow a clear visualization of the particles (see figure \ref{fig:MMF150H_particles_Haloes}). 
This is the simulation we use when (in the following sections) we refer to dark matter particles, 
unless we state something different.
\item The intermediate resolution ($n_{part} = 256^3$) is used to compute 
density fields in order to take full advantage of the spatial information with the
computing resources available.
\item The high resolution version ($n_{part} = 512^3$) is used to produce the HOP and FracHOP halo 
and subhalo catalogues (see sect.~\ref{sec:haloes}). 
\end{itemize}

\subsection{The density field}
The output of the N-body simulation consist of a discrete set of particles. This is 
translated into a continuum volume-filling density field sampled on a regular three 
dimensional grid. Crucial for the ability of the Multiscale Morphology Filter 
to identify anisotropic features such as filaments and walls is the use of a 
morphologically unbiased and optimized continuous density field retaining all 
features visible in a discrete galaxy or particle distribution. 

We therefore use the Delaunay Tessellation Field Estimator (DTFE), introduced by \citet{Schaapwey00} 
(see \cite{Schaap07,WeySchaap09} for extensive descriptions), to reconstruct 
the underlying density field. It uses the Voronoi and Delaunay tessellation of the particle 
distribution to obtain a optimal local density estimate and subsequently interpolate these 
values linearly over the simulation volume. For our purpose of detecting weblike features the 
DTFE method has several important characteristics:

\begin{enumerate}
\item[$\bullet$] The self-adaptive nature of the Delaunay tessellation, and resulting sensitivity to 
all levels of substructure present in the particle distribution, makes it highly suited 
for a multi-resolution analysis of the hierarchically evolved large scale matter distribution.
\item[$\bullet$] The Delaunay tessellation follows the intrinsic anisotropies of the spatial matter 
distribution, resulting in a density field reconstruction which accurately traces and outlines 
the intricate and complex spatial patterns in the cosmic web. 
\item[$\bullet$] It does not introduce significant artificial features. The main artefacts concern 
diffuse low-density tetrahedral wings at the boundary between dense and underdense regions and 
a rather noisy reconstruction of the underdense regions. 
\end{enumerate}

\noindent Figure \ref{fig:MMF150H_particles_dtfe} shows a slice of 25 $\Mpch$ 
along the $z$ axis of a subbox of $50 \times 50 \Mpch$. The top left panel shows the
particle distribution and the top right panel shows the corresponding DTFE density field. A
zoomed region is shown in the bottom panel. DTFE yields a density field reconstruction 
in which nearly all structures present in the particle distribution are represented. It finds 
highly dense clumps of matter, as well as the tenuous voids. Also note the absence of artificial 
blobs in the inner regions of the voids. An additional virtue concerns the absence of a gridlike 
imprint in the DTFE density field, even while this is still visible in the particle distribution.

Accordingly, DTFE is used to process the particle distribution into a continuous density field 
$f_{\tiny{\textrm{DTFE}}}$ (top centre frame, fig.~\ref{fig:MMF150H_particles_dtfe}). 

\subsection{Haloes and Subhaloes}
\label{sec:haloes}
We used the HOP group finder \citep{Eisenstein98} for the identification of self-bound 
virialized haloes. Each of these haloes is considered a parent candidate which may
contain one or more subhaloes. HOP links particles by associating each particle to the densest 
of its n-closest neighbors, until it finally reaches the particle that is its own densest neighbor. 
For allocating particles to their halos we applied HOP with standard parameters $\delta_{out}=80$, 
$\delta_{saddle} =120$ and $\delta_{peak}=160$. Although the groups identified with HOP are nearly 
identical to those found with FoF \citep{Davis85}, they are less prone to involve artificial bridges 
between close groups. 

Galactic haloes embedded in groups of galaxies and clusters of galaxies are identified 
with subhaloes in the computer simulation. They are the bound groups that are clearly defined 
against the diffuse background particles that form the halo in which they are embedded.
In order to identify the bound subhaloes inside larger groups otherwise identified as 
single virialized objects, we use the FracHOP algorithm developed by \cite{Aragon07}. 
It is an elaboration of the HOP halo finder \cite{Eisenstein98} and exploits the topological 
properties of nested local maxima smoothed on a fixed scale. It starts by rerunning HOP,  
exclusively for the particles identified with parent halos in the first HOP halo identification 
step. To this end, it uses a Gaussian smoothed density field with a kernel size of $R_f=35 $h$^{-1}$ kpc, 
so that substructure on smaller scale is suppressed. The subhalo identification is performed without
running REGROUP, so that all particles are assigned to their local maximum in the smoothed 
density field. The center of mass of each of the resulting candidate subhaloes is determined 
iteratively, while unbound particles are removed. The end product is a listing of the subhaloes 
within the simulation box.

Figure \ref{fig:MMF150H_particles_Haloes} shows the distribution of particles in the
$128^3$ and $256^3$ simulations as well as the distribution of HOP and FracHop haloes
plotted on top of the density field. Both HOP and FracHOP haloes closely follow the
patterns of the Cosmic Web, revealing that haloes are fair tracers of the large scale 
matter distribution. The only difference between the two is that a given HOP halo can be 
formed by several FracHOP haloes. The distribution of haloes delineates the Cosmic Web in 
a more sparse and smooth way compared to the particles.

\vskip 0.5truecm
\section{Morphological Segmentation:\\ \ \ \ \ \ the MMF Formalism}
\label{sec:mmf}
The Multiscale Morphology Filter (MMF) is used for identifying and characterizing 
the different morphological elements of the large scale matter distribution in the Cosmic Web 
\cite{Aragon07b}. The formalism has been developed on the basis of visualization and feature 
extraction techniques in computer vision and medical research \citep{Florack92}. The technology, 
finding its origin in computer vision research, has been optimized within the context of feature 
detections in medical imaging. \cite{Frangi98} and \cite{Sato98} presented its operation for the 
specific situation of detecting the web of blood vessels in a medical image. 

The MMF morphological segmentation takes account of the multiscale nature of the matter distribution 
by means of a Scale Space analysis, looking for morphological structures of mathematically 
specified type in a multiscale, scale independent, manner. The Scale Space analysis presumes that 
the specific structural characteristic is quantified by some appropriate parameter. Examples are 
density, eccentricity, orientation and curvature. The MMF filters these data to produce a hierarchy of 
maps having different resolutions, and subsequently selects at each point the dominant parameter value 
from the hierarchy in order to construct a scale independent map. 

The MMF-based dissection and visualization of the cosmic web in its three basic components allows us to 
concentrate on the significant features of the cosmic matter distribution, and reach a level of 
abstraction by avoiding spurious details. For the visualization of the intricate filament-cluster 
network this is particularly useful. In this section we briefly summarize the steps involved in the 
morphological segmentation of the cosmic web obtained from the N-body cosmological simulation. 
A detailed step-by-step description of the MMF algorithm can be found in \cite{Aragon07b}.

\begin{table*}
\label{tab:chap3_eigen_ratios}
\centering
\begin{large}
\begin{tabular} {|l|c|l|}
\hline
\hline
&&\\
Structure & $\lambda$ ratios & $\quad$ $\lambda$ constraints \\
&&\\
\hline
&&\\
Cluster Node (blob) &  $\lambda_1 \simeq \lambda_2 \simeq \lambda_3$ & $\lambda_3 <0\,\,;\,\, \lambda_2 <0 \,\,;\,\, \lambda_1 <0 $  \\
Filament     &  $\lambda_1 \simeq \lambda_2 \gg    \lambda_3$ & $\lambda_3 <0 \,\,;\,\, \lambda_2 <0  $  \\
Sheet    &  $\lambda_1 \gg    \lambda_2 \simeq \lambda_3$ & $\lambda_3 <0 $    \\
&&\\
\hline
\hline
\end{tabular}
\end{large}
\vskip 0.25truecm
\caption{Morphology and Eigenvalue configuration. The eigenvalue 
         conditions specify clusters, filaments and walls, each having 
         a density higher than the background. A negative eigenvalue indicates that the 
         feature reaches a maximum along the corresponding direction (and vice versa), while 
         a small eigenvalue indicates a low rate of change of the field values in the corresponding 
         eigen-direction (and vice versa). This leads to the morphological relationships listed 
         in this table.}
\end{table*}

\subsection{Scale Space}
The DTFE density field $f_{\tiny{\textrm{DTFE}}}$ is the starting point of the morphological segmentation. 
The density field is smoothed over a range of scales by means of a hierarchy of spherically symmetric Gaussian 
filters $W_{\rm G}$ having different widths $R_n$. The $n^{th}$ level smoothed version of the DTFE reconstructed field 
$f_{\tiny{\textrm{DTFE}}}$ is assigned $f_n$, 
\begin{equation}
f_{\rm n}({\vec x}) =\, \int\,{\rm d}{\vec y}\,f_{\tiny{\textrm{DTFE}}}({\vec y})\,W_{\rm G}({\vec y},{\vec x})\nonumber
\end{equation}
where $W_{\rm G}$ denotes a Gaussian filter of width $R_n$: 
\begin{equation}
W_{\rm G}({\vec y},{\vec x})\, = \,{1 \over ({2 \pi} R^2)^{3/2}}\, \exp \left(- {|{\vec y}-{\vec x}|^2 \over 2 R_n^2}\right)\,.
\label{eq:filter}
\end{equation}

Scale Space itself is constructed by stacking these variously smoothed data sets, yielding the family $\Phi$ of smoothed density maps $f_n$:
\begin{equation}
\label{eq:scalespace}
\Phi\,=\,\bigcup_{levels \; n} f_n 
\end{equation}
A data point can be viewed at any of the scales where scaled data has been generated.  The crux of the concept of Scale Space is that the 
neighbourhood of a given point will look different at each scale.  There are potentially many ways of making a comparison of the scale 
dependence of local environment. We address the local ``shape'' of the density field. 

\subsection{Local Shape}
The local shape of the density field at any of the scales $R_n$ in the Scale Space representation of the density field can be quantified on 
the basis of the Hessian matrix, ${\tilde {\mathcal H}}_{ij}=\nabla_{ij} f_n({\bf x})$, 
\begin{eqnarray}
&&\frac{\partial^2}{\partial x_i \partial x_j} f_n({\vec x})\,=\,f_{\tiny{\textrm{DTFE}}}\,\otimes\,\frac{\partial^2}{\partial x_i \partial x_j} W_{\rm G}(R_{\rm n})\nonumber \\
&&= \int\,{\rm d}{\vec y}\,f({\vec y})\,\,\frac{(x_i-y_i)(x_j-y_j)-\delta_{ij}R_{\rm S}^2}{R_{\rm S}^4}\,W_{\rm G}({\vec y},{\vec x})
\end{eqnarray} 
where ${x_1,x_2,x_3}={x,y,z}$ and $\delta_{ij}$ is the Kronecker delta. In other words, at each level $n$ of the scale space representation 
the Hessian matrix is evaluated by means of a convolution with the second derivatives of the Gaussian filter, also known as the Marr 
(or, less appropriately, ``Mexican Hat'') Wavelet. In order to properly compare the values of the Hessian arising from the differently 
scaled variants of the data that make up the Scale Space, the Hessian is renormalized, $\tilde {\mathcal{H}}\,=\,R_{\rm S}^2 \,\mathcal{H}$, 
where $R_s$ is the filter width that has been used. 

The eigenvalues $\lambda_i$ of the Hessian matrix determine the local morphological signal, dictated by the local shape of the 
density distribution. A small eigenvalue indicates a low rate of change of the field values in the corresponding eigen-direction, and vice versa.
We denote these eigenvalues by $\lambda_{a}(\vec{x})$ and arrange them so that $ \lambda_1 \ge \lambda_2 \ge \lambda_3 $:
\begin{eqnarray}
\qquad \bigg\vert \; \frac{\partial^2 f_n({\vec x})}{\partial x_i \partial x_j}  - \lambda_a({\vec x})\; \delta_{ij} \; \bigg\vert  &=& 0,  \quad a = 1,2,3 \\
\mathrm{with} \quad && \lambda_1 > \lambda_2 >  \lambda_3 \nonumber
\end{eqnarray}
The $\lambda_{i}(\vec{x})$ are coordinate independent descriptors of the behaviour of the density field in the locality of the point $\vec{x}$ and can be combined to create a variety of morphological indicators. The criteria we used for identifying a local bloblike cluster, 
filamentary or sheetlike morphology are listed in table~2. Evidently, the eigenvalues corresponding to a 
blob (cluster) morphology are a subset of the eigenvalue subset related to filamentary morphologies. In turn, the eigenvalue set of 
the latter is a subset of the wall eigenvalues. 

\subsection{Multiscale Structure Identification}
In practice, we are interested in the local morphology as a function of scale. In order to establish how it 
changes with scale, we evaluate the eigenvalues and eigenvectors of the renormalised Hessian $\tilde {\mathcal{H}}$ 
of each dataset in the Scale Space $\Phi$. 

Since we are looking for three distinct structural morphologies - cluster blobs, walls and filaments - the practical 
implementation of the segmentation consists of a sequence of three stages. Because curvature components are 
used as structural indicators, the blobs need to be eliminated before looking for filaments, after which the filaments 
have to be eliminated before looking for walls. This results in the MMF procedure following the sequence 
``clusters $\rightarrow$ filaments $\rightarrow$ walls''. At each of these three steps, the regions and scales are 
identified at which the local matter distribution follows the corresponding eigenvalue signature. 

In practice, the MMF defines a set of morphology masks, morphology response filters and morphology filters for each 
of the three different morphological components: clusters, filaments and walls. Their form is dictated by the particular 
morphological feature they seek  to extract, via the eigenvalues at each level in scale space and the criteria for each of 
the corresponding morphologies (table~2). The local value of the morphology response depends on the 
local shape and spatial coherence of the density field. The morphology signal $\Psi({\vec x})$ at each location is then defined to 
be the one with the maximum response across the full range of smoothing scales. Formally, we denote $\Psi$ by the name of 
\textit{Scale-Space Nap Stack}.

\begin{figure*}
  \centering
  \includegraphics[width=0.3\textwidth,angle=0.0]{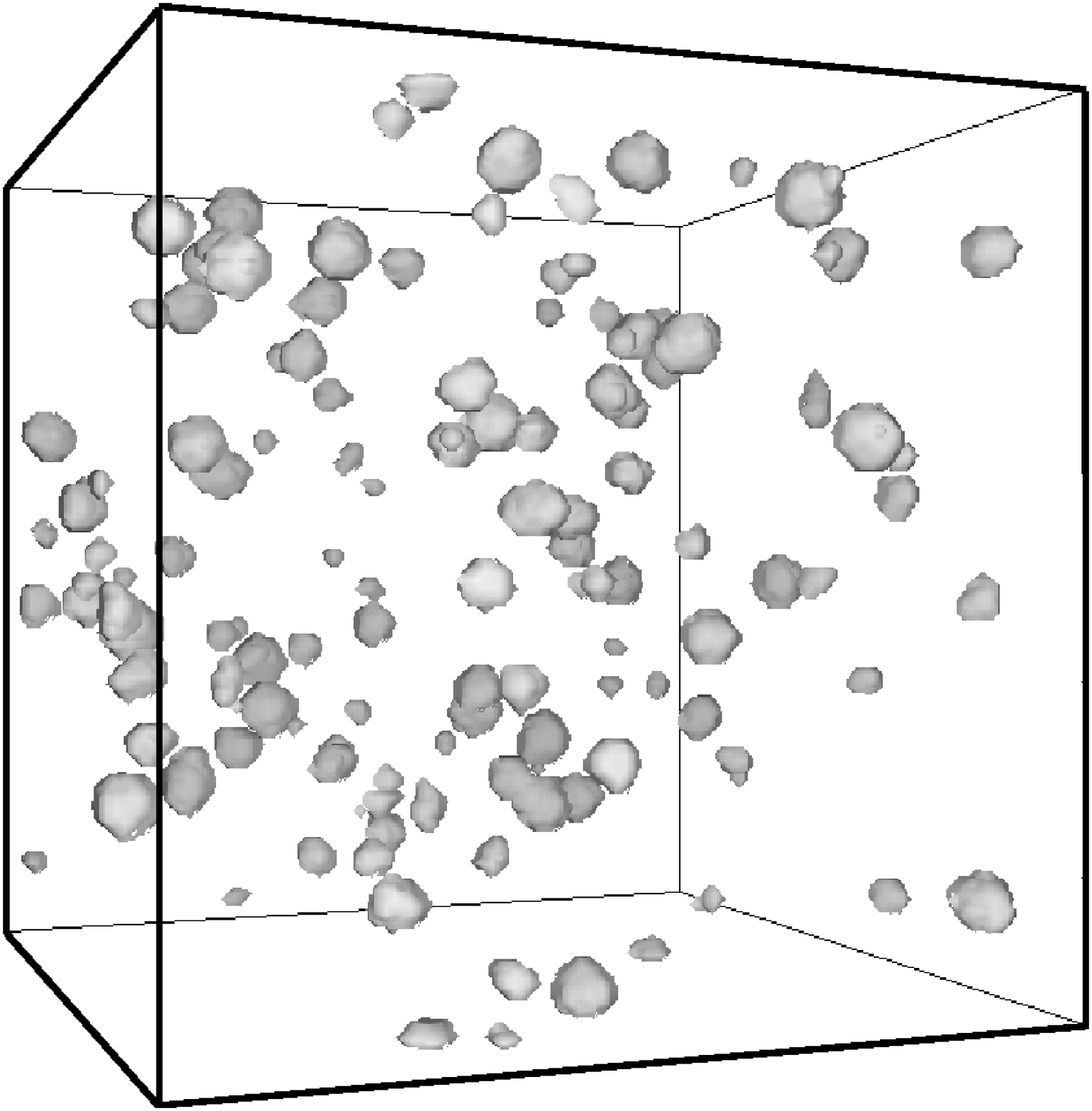}
  \includegraphics[width=0.3\textwidth,angle=0.0]{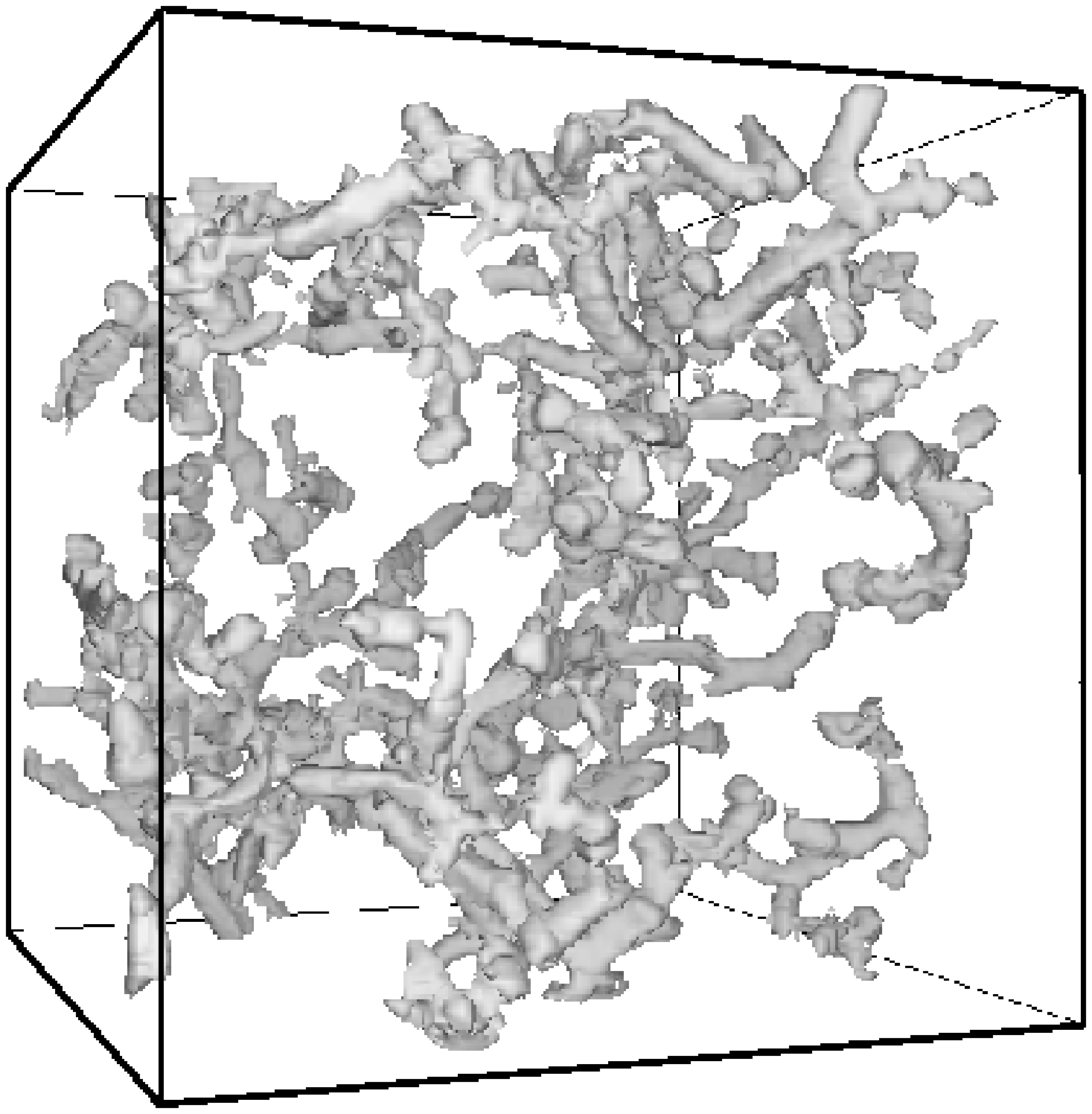}
  \includegraphics[width=0.3\textwidth,angle=0.0]{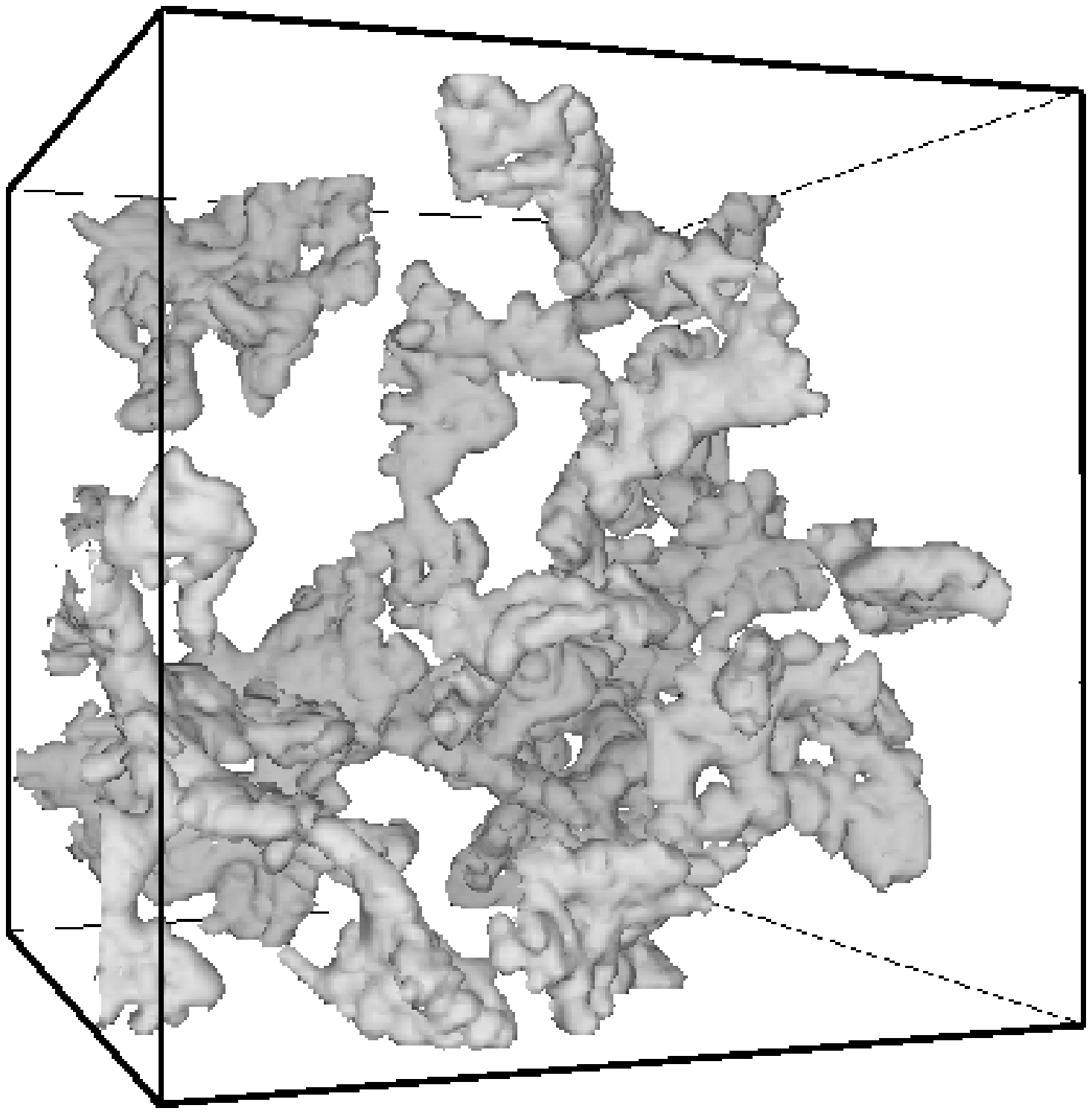}
  \includegraphics[width=0.32\textwidth,angle=0.0]{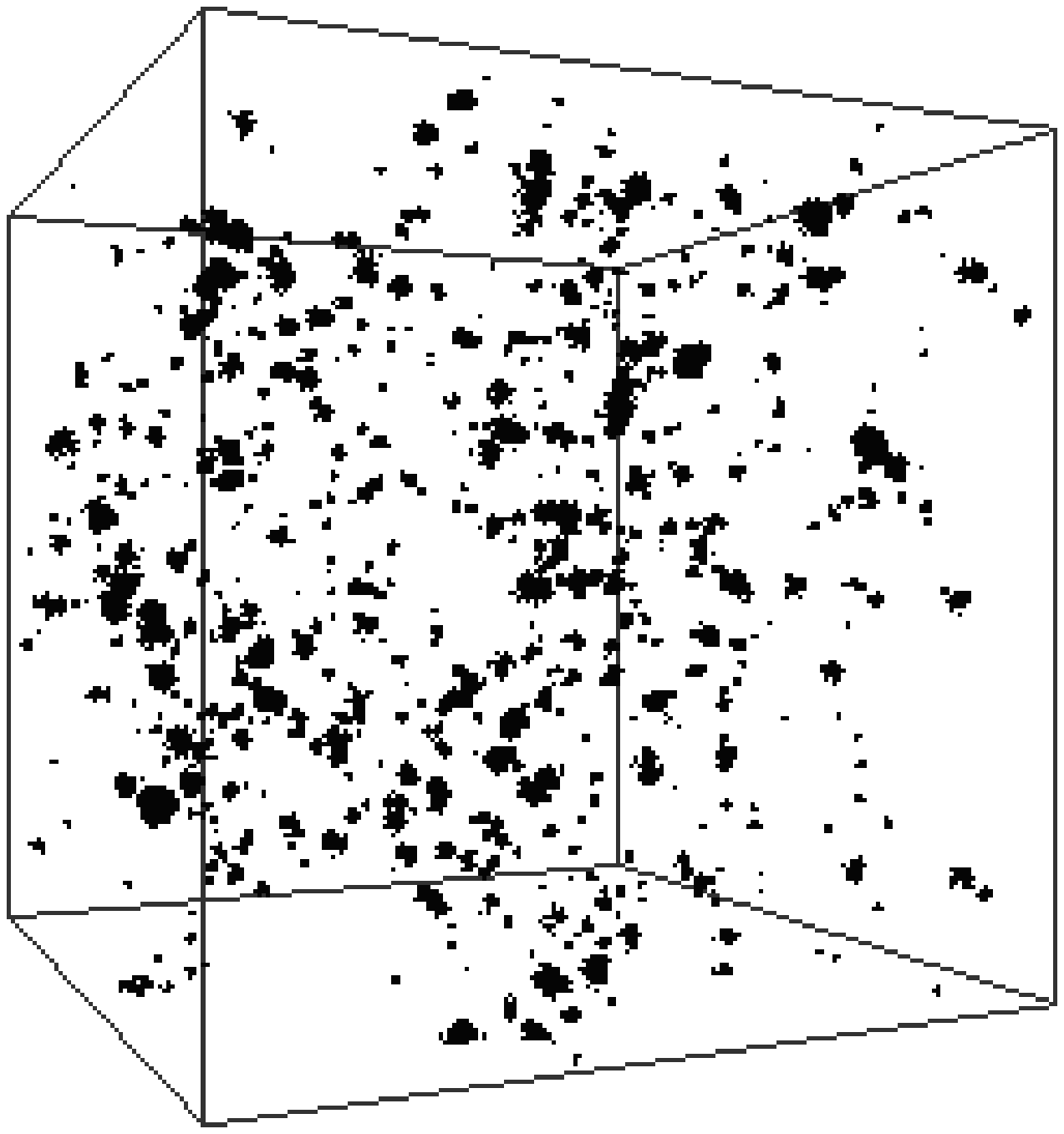}
  \includegraphics[width=0.32\textwidth,angle=0.0]{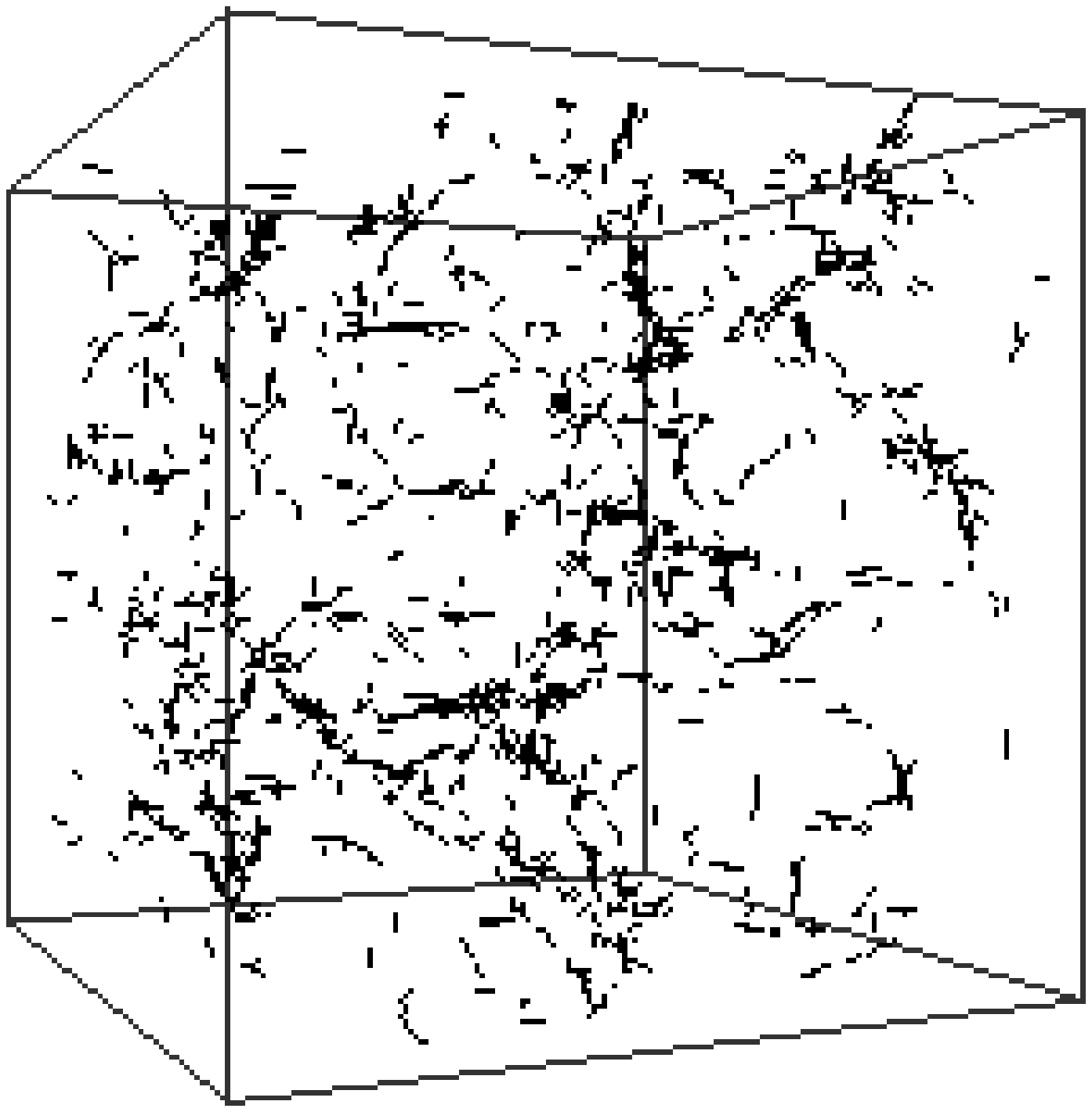}
  \includegraphics[width=0.32\textwidth,angle=0.0]{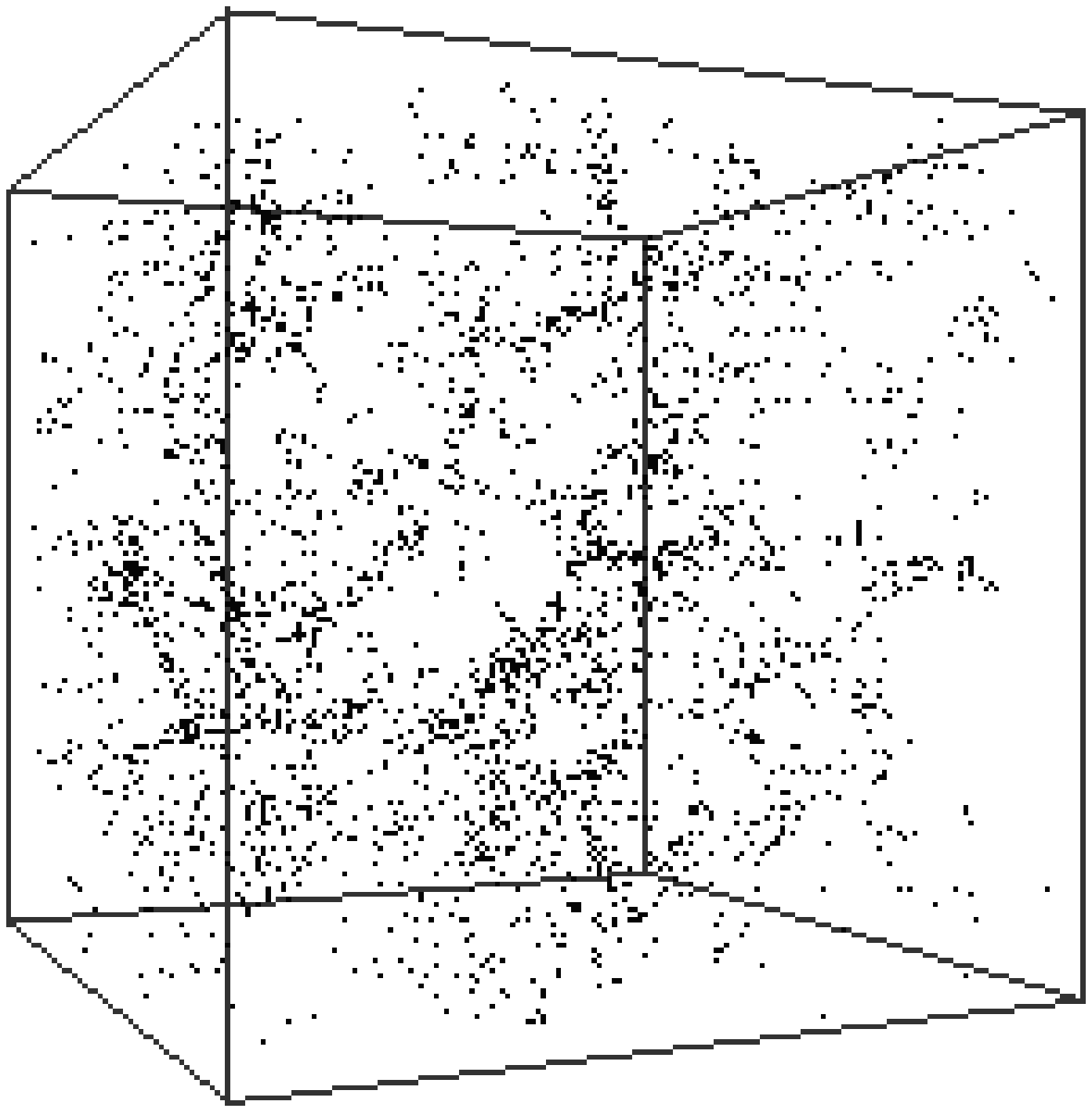}
    \caption{Top: Surfaces enclosing regions of space identified as clusters, filaments and walls 
      in the simulation box (left, middle and right panels). 
      Bottom: The particles enclosed by the surfaces in the top panels.
      Only the largest objects are shown for clarity. }
  \label{fig:MMF150H_MMF_blob}
\end{figure*} 

\subsection{Morphology Thresholds}
\label{sec:threshold}
The final step in the MMF feature identification concerns the removal of noisy structures. To this end, MMF 
invokes global morphology thresholds in order to separate the {\it texture noise} from genuine structures 
\citep[see][]{Aragon07b}. Regions with a morphology response $\Psi({\vec x})$ lower than the global threshold 
$\tau$ are omitted from the list of identified structures. 

The value of the thresholds $\tau_B$, $\tau_f$ and $\tau_w$ for clusters, filaments and walls is determined on the basis 
of the measured dependence of the structure detection rate as a function of the value of the morphology signal $\Psi$. 
All clusters with a morphology value less than the threshold $\tau_B$ are considered to be small insignificant blobs. 
The threshold is chosen such that these are erased, but not yet the the large gravitationally bound clumps. 
In the case of filaments and walls, the threshold value is determined on the basis of the percolation properties of the 
network of filaments and walls. The threshold values $\tau_f$ and $\tau_w$ are defined as the morphology 
signal value $\Psi$ for which the population of filaments and walls reaches its maximum number: at lower values the 
filaments and walls start to percolate.

In a distribution where all filaments or walls have similar properties such as contrast and physical extent 
this is the perfect choice. In the real Universe, however, there is a large variation in the contrast and 
size of filaments and walls. The same global criteria are therefore applied to faint as well as prominent 
structures. As a result, there is a systematic inclusion of low density regions forming the boundary 
of faint structures. While most mass is concentrated in high density regions, most of the volume of space 
concerns low-density regions. As a result, MMF has some bias towards low density structures. This might be 
alleviated by the use of more restrictive threshold values. However, this would imply the loss of very 
faint structures. A more preferrable but as yet not practical approach would be the use of a local threshold 
value which would account for the significance of features within the environmental context. 

\subsection{MMF Product}
The end product of the MMF procedure is a map segmented in clusters, filaments and walls (fig.~\ref{fig:MMF150H_MMF_blob}). 
These have been identified as the most outstanding structures and vary in scale over the full range of scales represented 
in Scale Space (eqn.~\ref{eq:scalespace}). Following the thresholding of the Scale-Space Map Stack $\Psi$ on the basis of 
cosmological and astrophysical considerations, we are left with the \textit{Object Map} $\mathcal{O}$. For each of the 
different morphologies - clusters, filaments and walls - these consist of the physically recognizable objects in the 
Cosmic Web. 

\bigskip
\section{Morphological Segmentation:\\ \ \ \ \ \ \ Cosmic Web Components}
\label{sec:webcomp}
Figure \ref{fig:cosmic_web_surfaces} shows the morphological segmentation of 
the 150$\Mpch$ simulation (150$_{\textrm{Low}}$ in table~2 obtained with the MMF. 
This figure illustrates  the large scale distribution of matter as an interconnected  network of 
filaments (dark gray) defining the boundaries of walls (light gray), and the clusters (black) located 
at the intersections of the network. The spatial distribution for each of the individual morphological components 
is shown in fig.~\ref{fig:MMF150H_MMF_blob}, by means of isodensity surfaces (top column) and 
by means of the particles enclosed by these surfaces (bottom column). 
 
For clarity fig.~\ref{fig:cosmic_web_surfaces} and fig.~\ref{fig:MMF150H_MMF_blob} only show the largest 
structures. Including all the objects would quickly have produced an image saturated with walls and filaments. By 
restricting the number of structures included, it is easier to identify the individual components of the 
Cosmic Web. Each morphological component is well differentiated and occupies, by construction, mutually 
exclusive regions. The variety of sizes of the clusters is a nice illustration of the ability of the MMF to 
identify structures at different scales (also see sec.~\ref{sec:clusters}).

Careful inspection of figs.~\ref{fig:cosmic_web_surfaces} and ~\ref{fig:MMF150H_MMF_blob} reveals 
the close physical affiliation of the different morphological components. Cluster blobs are located at the 
tips of filaments, and filaments tend to be found at the boundaries of walls. This confirms theoretical 
expectations \citep{Zeldovich70,Bondweb96}.

Following the MMF segmentation of the matter distribution acccording to their intrinsic morphology 
and scale, it is straightforward to compute the global properties of each morphological component, 
such as mass and volume content. For other properties, such as the length and density profiles of 
filaments, additional post-processing steps are necessary. These analysis procedures will be 
described in the next sections. First, we present a qualitative and illustrative impression of 
each of the main morphologies, starting with clusters and followed by filaments and walls. 

\begin{figure*}
  \centering
  \includegraphics[width=0.75\textwidth,angle=0.0]{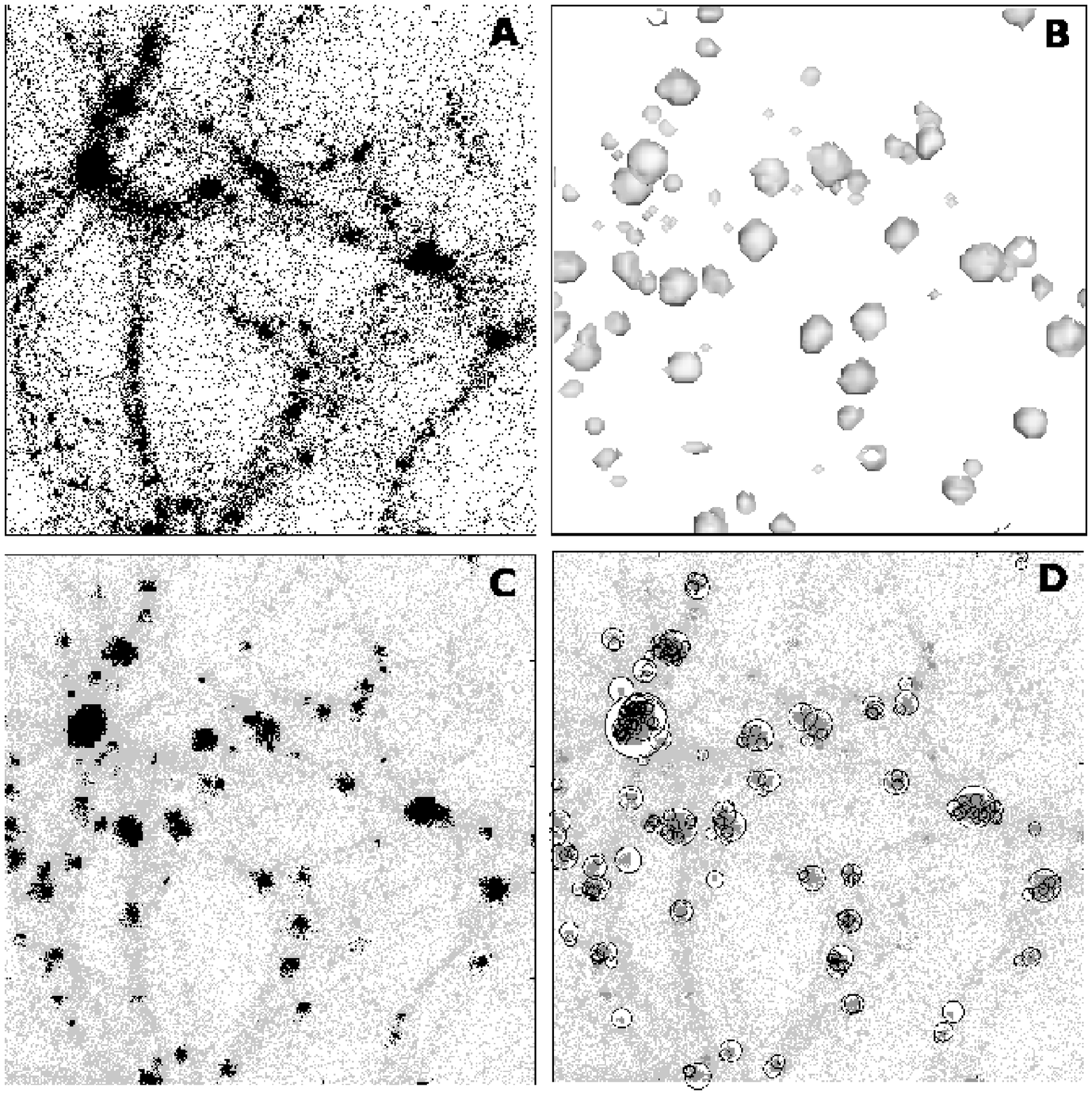}
    \caption{MMF Cluster \& Halo Identifications. A).- All particles inside a sub-box of the simulation. 
    B).- Surfaces enclosing regions identified by the MMF as clusters. Note that projection effects may distort 
    the real size of the surfaces enclosing clusters. C).- Particles located inside MMF clusters (dark grey). The rest of the 
    particles are shown in light grey color. D).- Halos identified with HOP that have their center of mass
    inside regions identified as clusters by the MMF. The MMF manages to identify the dense clusters
    at their characteristic scale. This may be appreciated from panels C and D. The match is reasonably good, 
    (although not perfect due to the intrinsic differences between HOP and the MMF methods). }
  \label{fig:MMF150H_MMF_blob_part_halo}
  \centering
  \mbox{\hskip -0.75truecm\includegraphics[width=0.42\textwidth,angle=0.0]{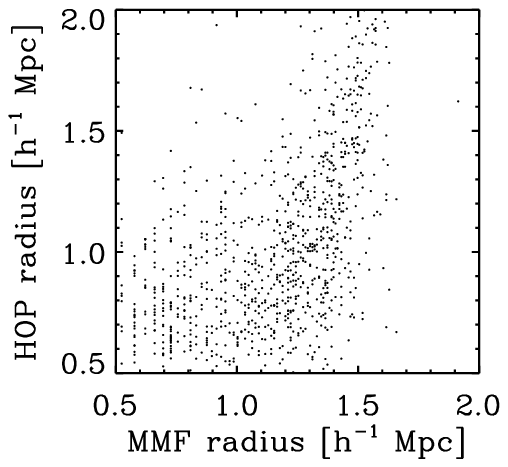}}
  \mbox{\hskip -0.75truecm\includegraphics[width=0.42\textwidth,angle=0.0]{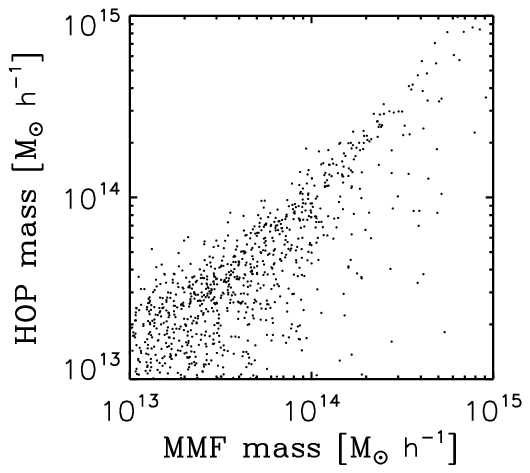}}
    \caption{Halo Identification Comparison. Scatter plot of the radius (left) and mass (right) of haloes identified with 
             HOP and the MMF. The radius of the HOP haloes corresponds to the virial radius}
  \label{fig:MMF150H_MMF_HOP_radius_mass}
\end{figure*} 
\begin{figure*}
  \centering
  \includegraphics[width=0.95\textwidth,angle=0.0]{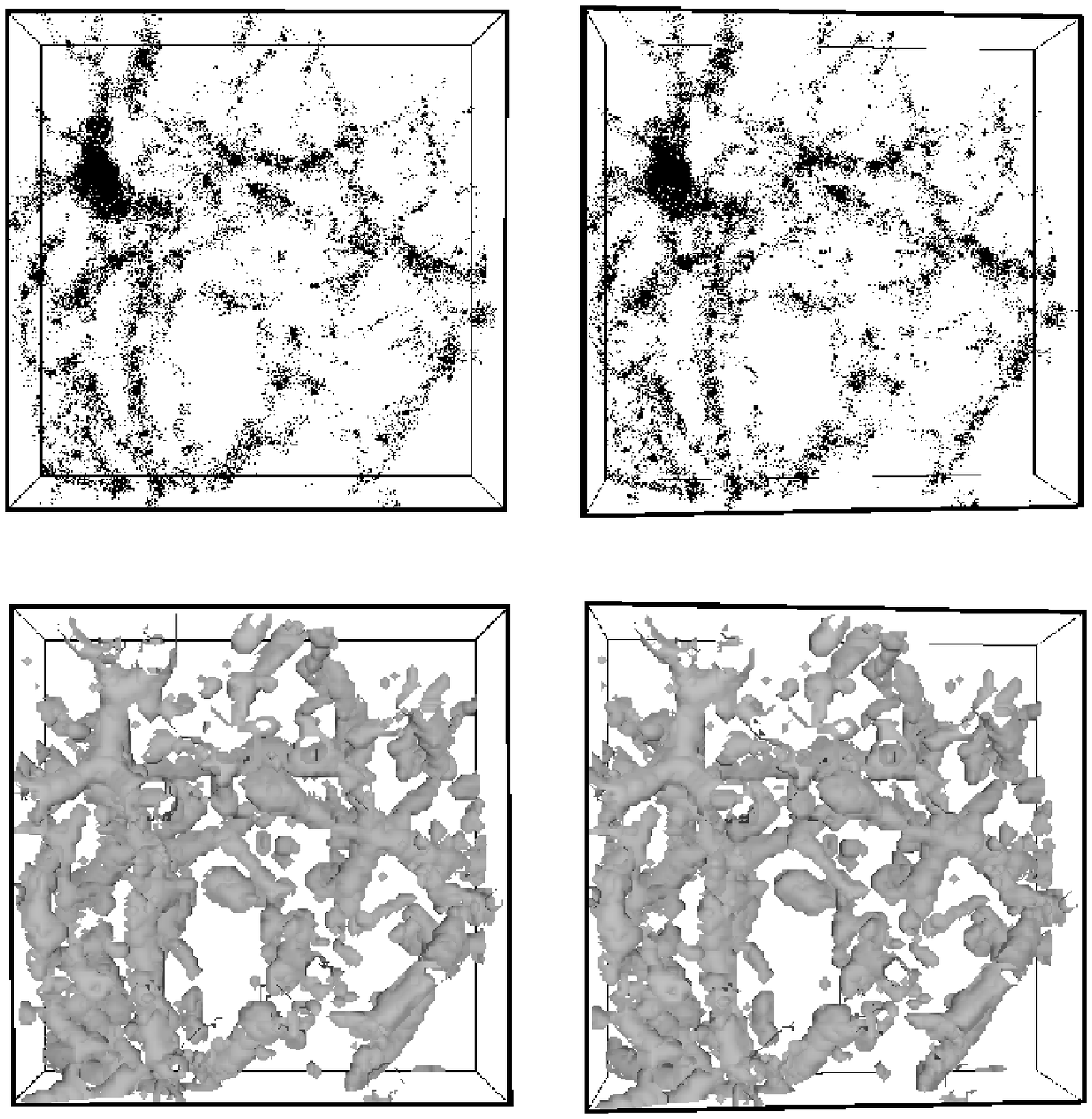}
    \caption{MMF Filaments. 3D stereoscopic view (Cross eyed) of the isosurfaces enclosing filaments (bottom panels) and the
    	enclosed particle distribution (top panels). The box corresponds to the zoomed region shown in 
	Figure \ref{fig:MMF150H_particles_Haloes}}
  \label{fig:MMF150H_MMF_fila}
\end{figure*} 

\begin{figure*}
  \centering
  \includegraphics[width=0.95\textwidth,angle=0.0]{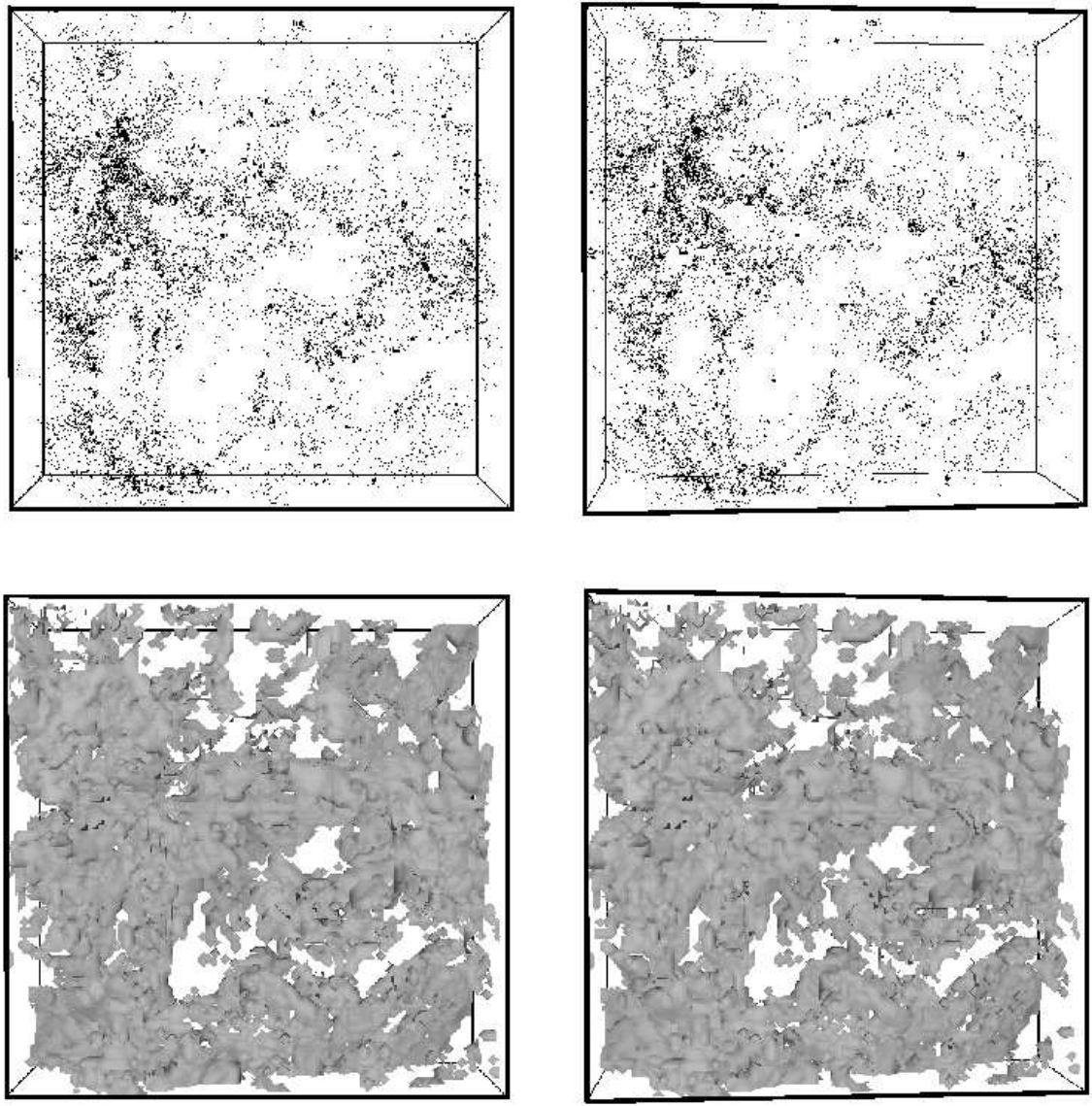}
    \caption{MMF Walls. 3D stereoscopic view (Cross eyed) of the isosurfaces enclosing walls (bottom panels) and the
    	enclosed particle distribution (top panels). The box corresponds to the zoomed region shown in 
	Figure \ref{fig:MMF150H_particles_Haloes}}
  \label{fig:MMF150H_MMF_wall}
\end{figure*}

\subsection{Clusters}
\label{sec:clusters}

A more detailed view of the cluster distribution is presented in figure \ref{fig:MMF150H_MMF_blob_part_halo}. We 
illustrate the cluster distribution by means of four panels:

\begin{itemize}
\item[\textbf{A})] The particle distribution.
\item[\textbf{B})] Surfaces defining clusters identified with the MMF.
\item[\textbf{C})] Particles identified inside MMF clusters.
\item[\textbf{D})] HOP haloes corresponding to the MMF clusters.
\end{itemize}

\noindent The size of the objects seen in the particle distribution as well as the HOP haloes
is related to the size of the clusters identified with the MMF. The match is not perfect,
as one may expect due to the intrinsic differences between HOP and the MMF. This is
illustrated in figure \ref{fig:MMF150H_MMF_HOP_radius_mass}, where we compare the radius and mass 
of clusters identified with HOP and the MMF. 

The radius of HOP clusters is defined as the distance from the center of mass to outermost particle. 
For the MMF clusters, we computed the radius from 

\begin{equation}
R = \left ( \frac{3}{4 \pi} V_{blob} \right )^{1/3}. 
\end{equation}

\noindent where $V_{blob}$ is the volume of all pixels defining an individual blob (cluster).
The masses of HOP and MMF clusters are well correlated. This is not surprising, since most of the 
mass of the cluster is located in the dense inner regions of the cluster. On the other hand, the scatter between 
radius of HOP and MMF clusters is large. This is a result of the way in which the radius is estimated. The 
distance of the most distant particle from the center of mass is sensitive to small fluctuations in the 
perifery of the clusters. Resolution effects also influence the estimated radius of MMF clusters, since the 
density field grid size is large compared to the radius of the smallest clusters. 

\begin{table*}
\begin{center}
\begin{large}
\begin{tabular} {l r r r r }
               &  Clusters &  Filaments &  Walls &  Voids\\
\hline
\hline
Volume filling   (\%)      & 0.4  & 8.8 & 4.9  & 85.9\\
Mass content     (\%)      & 28.1  & 39.2 & 5.5  & 27.2\\
Mean overdensity           & 73.0  & 4.5 & 1.1  & 0.3 \\
Median overdensity         & 11.5  & 1.7 & 0.9  & 0.3 \\
Standard deviation         & 58.8 & 11.4 & 2.61 & 0.52 \\
Kurtosis                   & 58.7  & 44.8 & 160.5 & 142.1 \\                
\hline
\end{tabular}
\end{large}
\caption{Inventory of the Cosmic Web. Listed are volume, mass content and a few statistical characteristics 
         of the density distribution of the individual structural morphologies. 
         Mean, median, standard deviation and kurtosis are computed from the distribution of the overdensity 
         $1+\delta = \rho / \overline{\rho}$. }
\end{center}
\label{tab:cosmic_web_inventory}
\end{table*}
The third major effect which influences the radius estimate of massive clusters is the often substantial 
intrinsic elongation of clusters. By virtue of the MMF formalism, clusters identified with the MMF tend to be  
spherically symmetric. However, in general the shape of virialized clusters tend to depart from 
sphericity, which can be most clearly observed in computer simulations \citep[see e.g][]{Araya09}. Massive clusters 
are often highly elongated in the direction of the filaments connected to them, and along which most merging 
clumps are moving in \citep{Haarlem93}. In fact, the infall of matter along the filamentary transport channels 
amplifies the elongation and alignments of the clusters with respect to the filaments and neighbouring 
clusters \citep{Haarlem93}. The strongest contribution to this effect is that by the merging of two or 
more clusters, which shares the highly anisotropic nature of the more gradual accretion of most of the 
matter. 

Once all clusters are identified, the corresponding cluster particles are removed from the particle 
sample. The cluster-free particle distribution is subsequently analyzed for its filament population.

\subsection{Filaments}
On the basis of the ``cluster-free'' particle distribution, we compute the DTFE density field. This 
density field only contains structures associated with filaments and walls, and is used as input for 
the identification of filaments in the MMF pipeline. 

The 3D visualization allows us to appreciate the three-dimensional nature of the filaments, in particular 
also highlighting their connectivity. Figure \ref{fig:MMF150H_MMF_fila} shows a 3D view of filaments inside 
a subbox of 50 $\Mpch$ of side. The isosurfaces enclosing regions of space identified as filaments 
are shown in the bottom panels. The particles they contain are shown in the top panels. Note that the regions 
where clusters are located (e.g. the large cluster at the top left corner) are also covered by the 
filament mask, but that we explicitly exclude these regions from the filament mask.  

Note that there are some small particles and clumps, which appear to be mostly isolated and bear no 
relation to the surrounding structures. These correspond to regions that have a spurious filamentary nature, 
and are near the resolution and filament detection limit of MMF. We discard these spurious identification 
from the filament mask. 

\begin{figure}
  \centering
  \includegraphics[width=0.35\textwidth,angle=0.0]{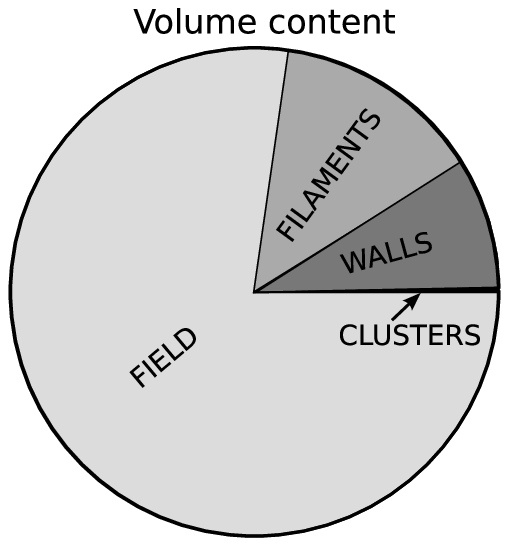}
  \vskip 0.5truecm 
  \includegraphics[width=0.35\textwidth,angle=0.0]{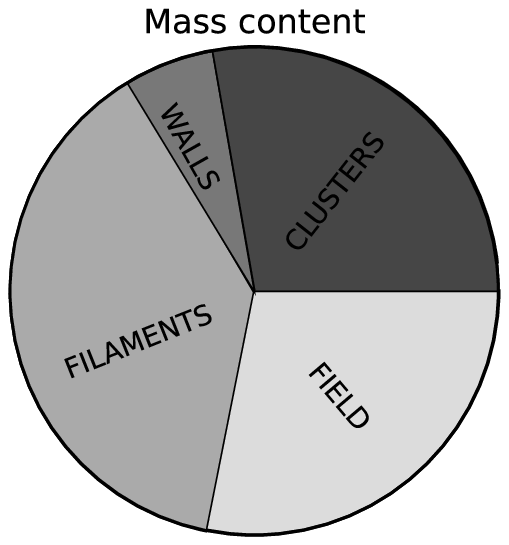}
    \caption{Pie diagram showing an inventory of the Cosmic Web in terms of volume (left) and mass (right). We 
    distinguish clusters, filaments, walls and void regions (or ``field''). }
  \label{fig:cosmic_web_pie_diagram}
\end{figure}

Filaments identified with the MMF are complex objects that pervade throughout the sample volume, connecting 
each of its regions. MMF follows the intrinsic scale of the real filaments, which feature a large range of 
lengths and widths. The corresponding filament particle distribution is far from homogeneous: along the 
filaments we find dense haloes as well as a pervasive medium of diffuse particles. It is clear that a description 
of such complex systems requires advanced methods of analysis which are sensitive to the anisotropic and 
multiscale nature of the matter distribution. The MMF is an example of such a specifically designed instrument. 

\subsection{Walls}
After removing the particles located inside clusters and filaments we proceed to the last step in the morphological 
segmentation, the identification of walls in the density field. Walls are the most tenuous coherent structures in the 
large-scale universe. Their identification poses a major challenge \citep{Shandarin04}, and their tenuous nature 
and complex topology and shapes makes them the most difficult morphology to characterize. 

Figure \ref{fig:MMF150H_MMF_wall} shows a 3D view of the walls inside a volume of 50$\Mpch$ size. The top panel 
shows the particles located within the walls, with the bottom panels depicting the corresponding walls by means of 
isosurfaces. The DTFE density field on the basis of the cluster and filament free particle sample contains only 
features of a planar nature, although it also contains various noisy features that cannot be clearly identified with 
any of the three basic morphologies. 

In general, walls identified with the MMF are far from smooth planar objects. They tend to have crumpled shapes with 
no obvious topology. Because they are multiply connected objects, it is virtually impossible to isolate individual 
walls from the complex web of walls. Evidently, MMF identifies the walls succesfully. However, as a result of their 
complex nature, we restrict ourselves to assessing their basic properties. 


\section{Cosmic Web Inventory}
\label{sec:webmass}
\subsection{Mass and volume content}
\label{sec:mass_and_volume_content}
In order to understand the role of clusters, filaments and walls in the shaping of the
cosmic web it is crucial to determine their relative abundances in terms of volume and
mass. Such quantities may provide rough estimates of the dominance of one morphology with
respect to the other in shaping the cosmic web and driving its overall dynamics.

Table~3 lists some basic characteristics for each of the 
morphological elements. These include an inventory of the Cosmic Web in terms of volume 
and mass content. The mass content of a particular morphology is measured by adding the 
total mass of the particles enclosed within the boundaries of that morphology. The fraction 
of the occupied volume is determined by adding the volume of all voxels enclosed within 
the morphological boundaries.  

The resulting cosmic inventory is summarized in the pie diagrams of fig.~\ref{fig:cosmic_web_pie_diagram}. 
The stark contrast between the volume and the mass share of the clusters, filaments and sheets is 
a direct manifestation of the large density differences between the different morphologies and a direct 
indication of the dynamical importance of these elements. The density contrast differences are also 
an indication for the different evolutionary stages in which they reside, as gravitational collapse 
proceeds faster as we go from $\;$ walls $\rightarrow$ filaments $\rightarrow$ clusters. 

\begin{figure}
  \centering
  \includegraphics[width=0.5\textwidth,angle=0.0]{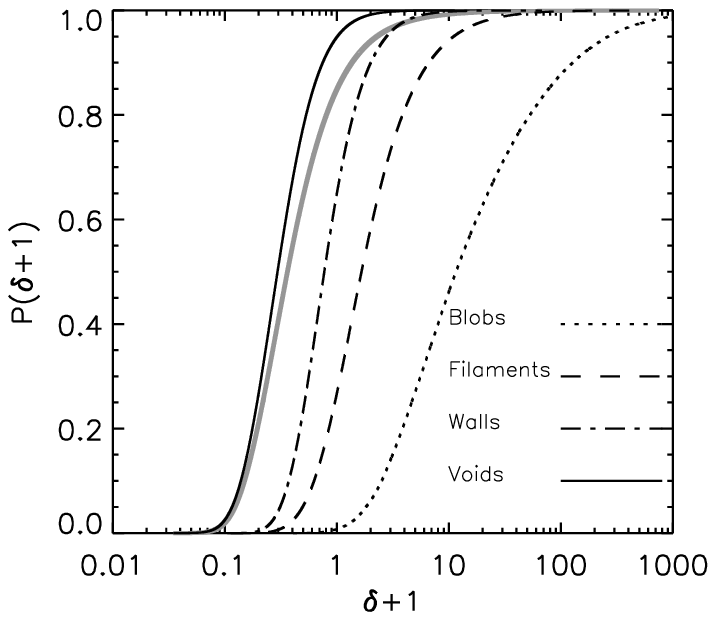}
  \includegraphics[width=0.5\textwidth,angle=0.0]{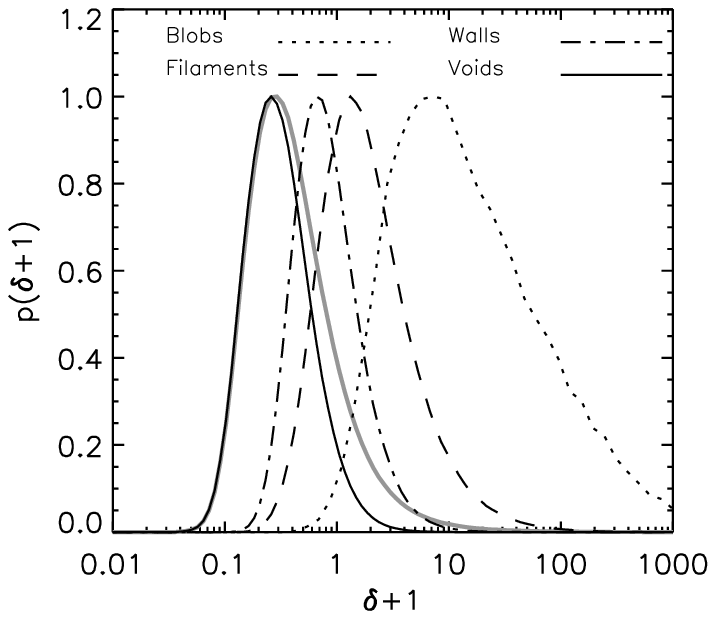}
    \caption{Cumulative (left) and density (right) probability distribution of the density contrast
             in clusters (blobs), filaments, walls and the void regions (dotted, dashed, dotted-dashed and solid respectively). 
             The thick grey lines indicates the distribution corresponding to all the volume.}
  \label{fig:density_cum}
\end{figure}

Not surprisingly, clusters occupy the smallest volume fraction in the cosmic web, occupying only 0.38 \%. 
Despite this, they also represent a major share of the cosmic mass: 28\% of the total mass resides within 
cluster regions. This not only makes them 
by far the densest objects of the Megaparsec universe, but also the dynamically dominant component of the 
Cosmic Web \citep[see e.g.][]{Bondweb96}. The largest fraction of the mass in the Universe, $\approx 39\%$, 
resides in filaments, which occupy around $10\%$ of the total volume. Although their density is 
lower than that of clusters, they represent the most salient component of the Cosmic Web via 
their function as the all-pervasive bridges between all structural features in the Megaparsec 
Universe. Walls contain a substantially smaller fraction of the mass, $\approx 5.5 \%$. They also 
occupy a relatively small volume, at $\approx 4.9\%$ even less than that occupied by filaments. It 
certainly means that walls have been relatively unimportant in the recent formation history of the cosmic web. 

It is instructive to compare the present-day morphological inventory with that in the primordial 
density field. On the basis of the (tidal) deformation tensor distribution in the primordial Gaussian 
field, \citep{Doroshkevich70} showed that in the \textit{linear regime} 92 \% of the mass will collapse
into walls, filaments or clusters. Filaments and walls would each take $42\%$ of the share, clusters 
$8\%$, while the remaining 8 $\%$ would correspond to underdense voids. While we may already expect that 
this direct link between primordial deformation tensor and morphology is too simplistic, it is the 
subsequent quasi-linear and nonlinear evolution which changes these numbers substantially. 

The MMF is a density field based criterion, and performs better as the density field becomes more prominent and 
non-linear: it selects only those regions with a clear morphology and contrast. Following the same deformation 
tensor criteria with respect to the primordial density field, \citep{Pogosyan98} showed that filaments will be much more 
prominent in the high density regions, which tend to develop faster in the subsequent nonlinear 
evolution. Walls are more biased to lower density regions, and at all times will therefore occur less 
prominent than filaments. The MMF sensitivities will therefore be naturally biased towards the 
filaments and clusters in the mass distribution. Moreover, MMF is less likely to properly 
identify the outer infall regions of clusters, filaments and walls and instead tends to relegate 
part of these to the field as they have not yet emerged as fully developed structures. Here, for 
simplicity, we identify these field regions with the larger low density voids.

\subsection{Density segregation of the cosmic web}
\label{sec:densseg}
The differences in mass and volume content derived for each morphology correspond to different 
density ranges. It is often assumed that these elements mark out a unique density regime, 
with no overlapping values. This assumption is the basis for the use of overdensity as one 
of the most widely used criteria to identify clusters \citep{Lacey94,Eke96} and filaments 
\citep{Shandarin04,Dolag06}. 

Figure \ref{fig:density_cum} shows the cumulative and probability distributions of the overdensity 
$\delta$ within the regions identified as clusters, filaments, walls and fied. The lower panel shows 
that each morphology occupies a characteristic range in density.

\begin{itemize}
\item Clusters are the densest objects, with a median overdensity of $\sim 11.5$ and a mean
overdensity of $\sim 73$. The range of overdensities extends to more than $\delta \sim 100$ and even 
$\delta \sim 1000$ within the large virialized clusters. 

\item Filaments and walls have medium overdensities, with mean overdensity of $\langle 1+\delta \rangle \sim 4.5$ for 
filaments and $\langle 1+\delta \rangle \sim 1.1$ for walls and a median overdensity of $~\sim 1.7$ and $\sim 0.9$ 
respectively. 

\item The field should be mostly identified with the most underdense void regions. On average, the void regions 
correspond to underdensities of $\delta \sim -0.7$. 
\end{itemize}

The density values for filaments and walls partially coincide with the density range expected for 
collapsed objects, since they concern values $\delta \geqslant 6$ at which spherical 
objects turn around into collapse. However, in particular for walls a major fraction of the enclosed 
space has a substantially lower density. To a large extent this concerns the lower density in the outer 
realms which surround the dense inner regions of clusters, filaments and walls. There is also a bias 
towards low densities in structures identified with the MMF as a result of the morphology threshold 
criteria used by MMF to separate real and noisy structures (see sect.~\ref{sec:threshold}). 

\begin{figure*}
  \centering
  \mbox{\hskip -0.85truecm\includegraphics[width=0.95\textwidth]{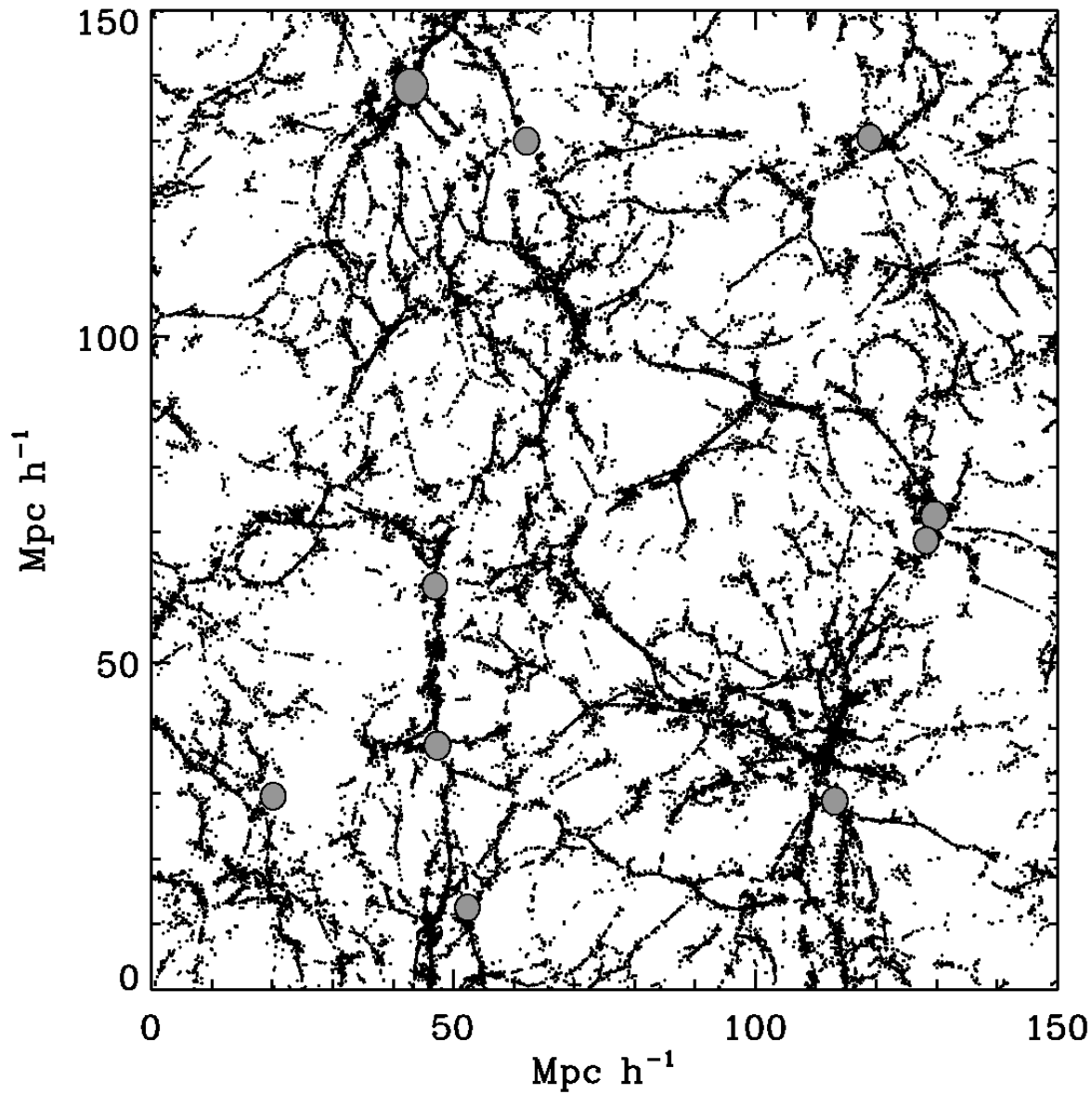}}
    \caption{Filamentary network in a slice of 20$\Mpch$. Back dots indicate dark matter particles in filaments
             after the compression algorithm. Gray circles indicate the location of
             clusters with $M \ge 10^{14} M_{\odot} h^{-1}$. The size is scaled proportional to their mass.}
  \label{fig:filamentary_network}
\end{figure*}

The considerable level of overlap in density between the various morphologies also means that a pure 
density criterion for structural identification does not provide an accurate description of 
reality. A morphological segmentation in terms of density alone would require at least a 
non-overlapping low-density tail. However, the fact that this does not seem to be the case 
implies there to be a substantial contamination with other morphological elements if one 
would resort to a pure density-based criterion: a (global) density threshold would be a poor 
discriminator of morphology. Even when each morphology can be associated with a specific density range, 
in general an additional, more sophisticated characterization is required.

The use of the MMF method to disentangle the cosmic web into its basic morphologies, independent of 
their density contrast, is clearly justified by the results presented in figure~\ref{fig:density_cum}.
However, we do have to take care of the fact that the MMF density estimates of filaments and 
walls tend to be systematically lower than the actual values (see sect.~\ref{sec:threshold}). 

\begin{figure*}
  \centering
  \includegraphics[width=0.42\textwidth,angle=0.0]{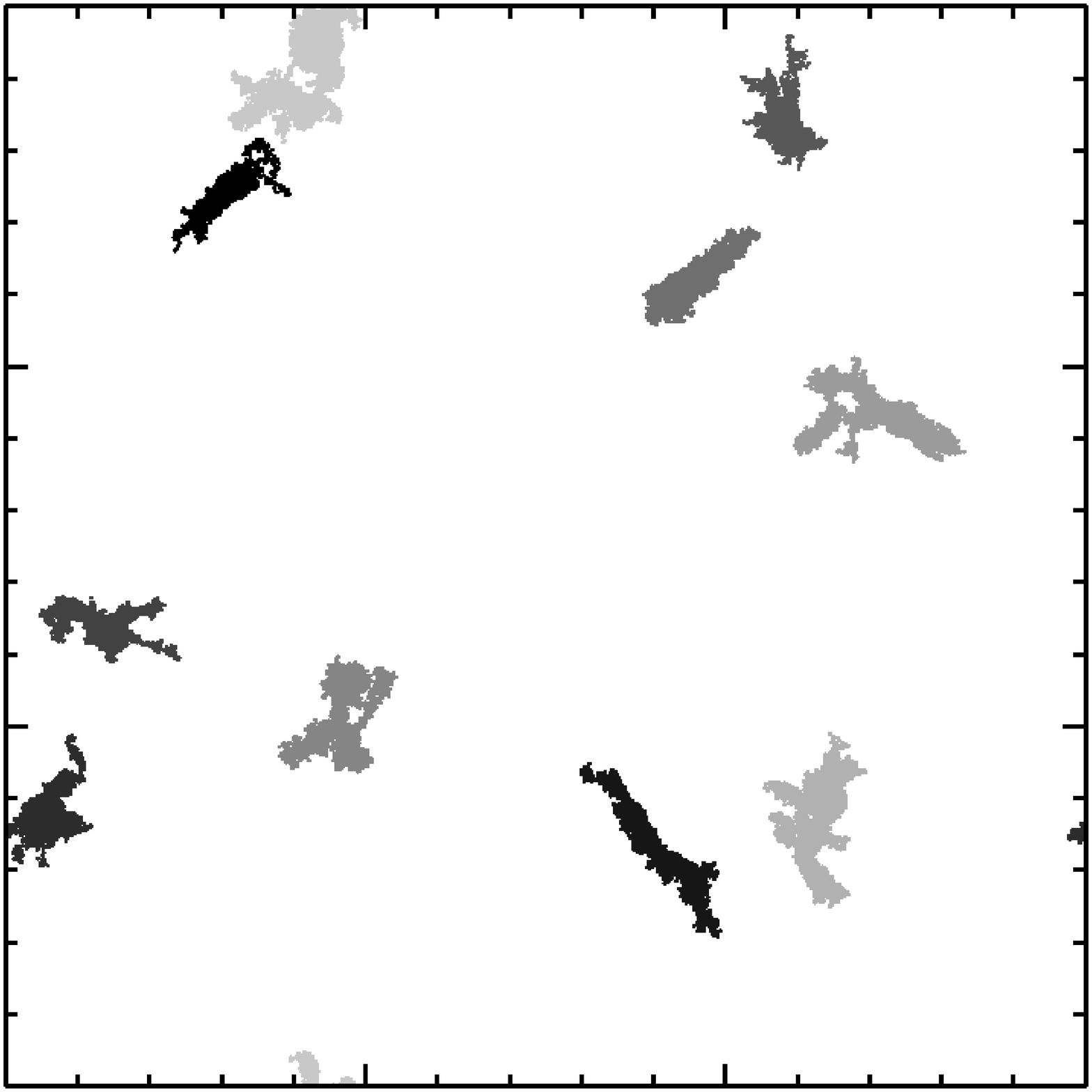}
  \includegraphics[width=0.42\textwidth,angle=0.0]{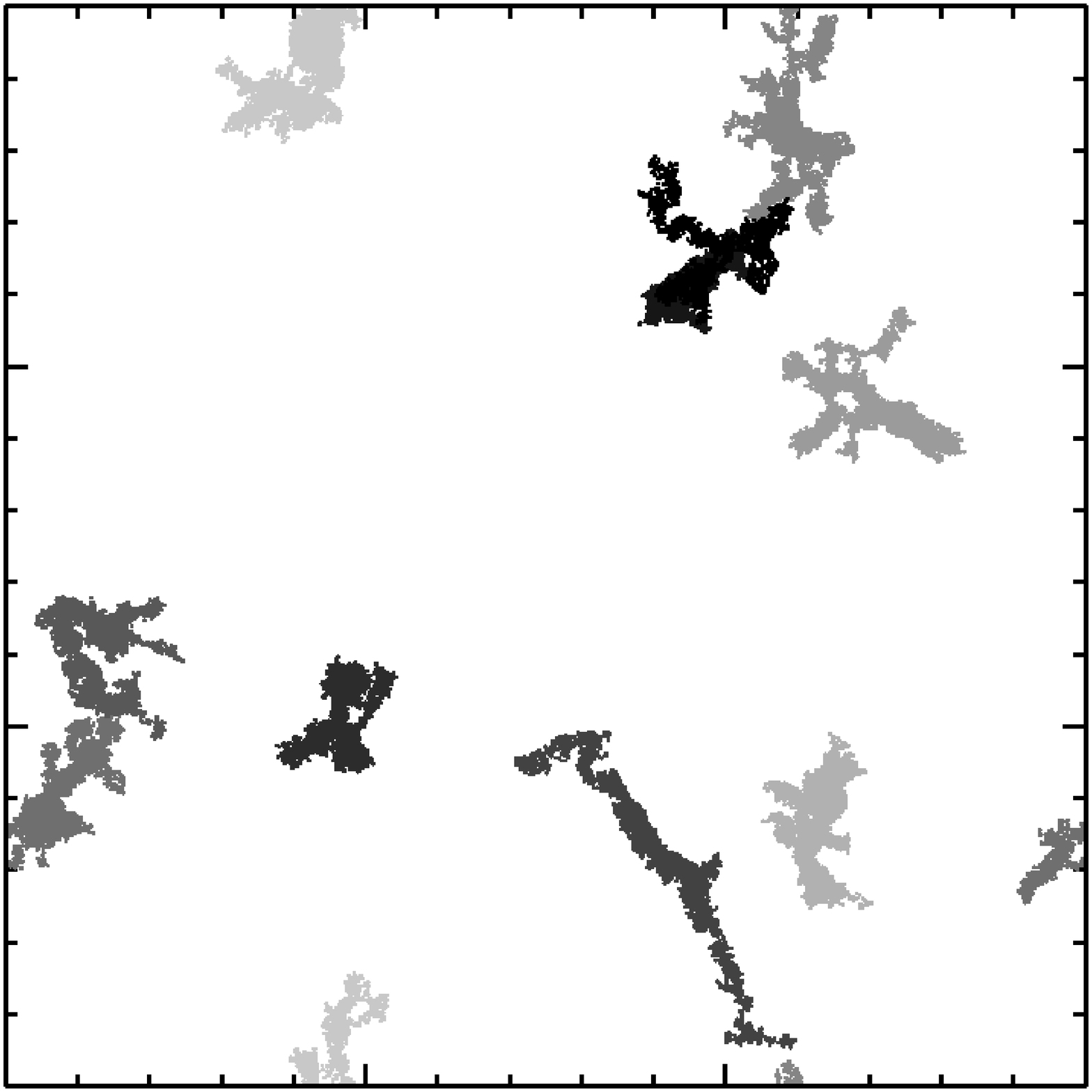}
  \includegraphics[width=0.42\textwidth,angle=0.0]{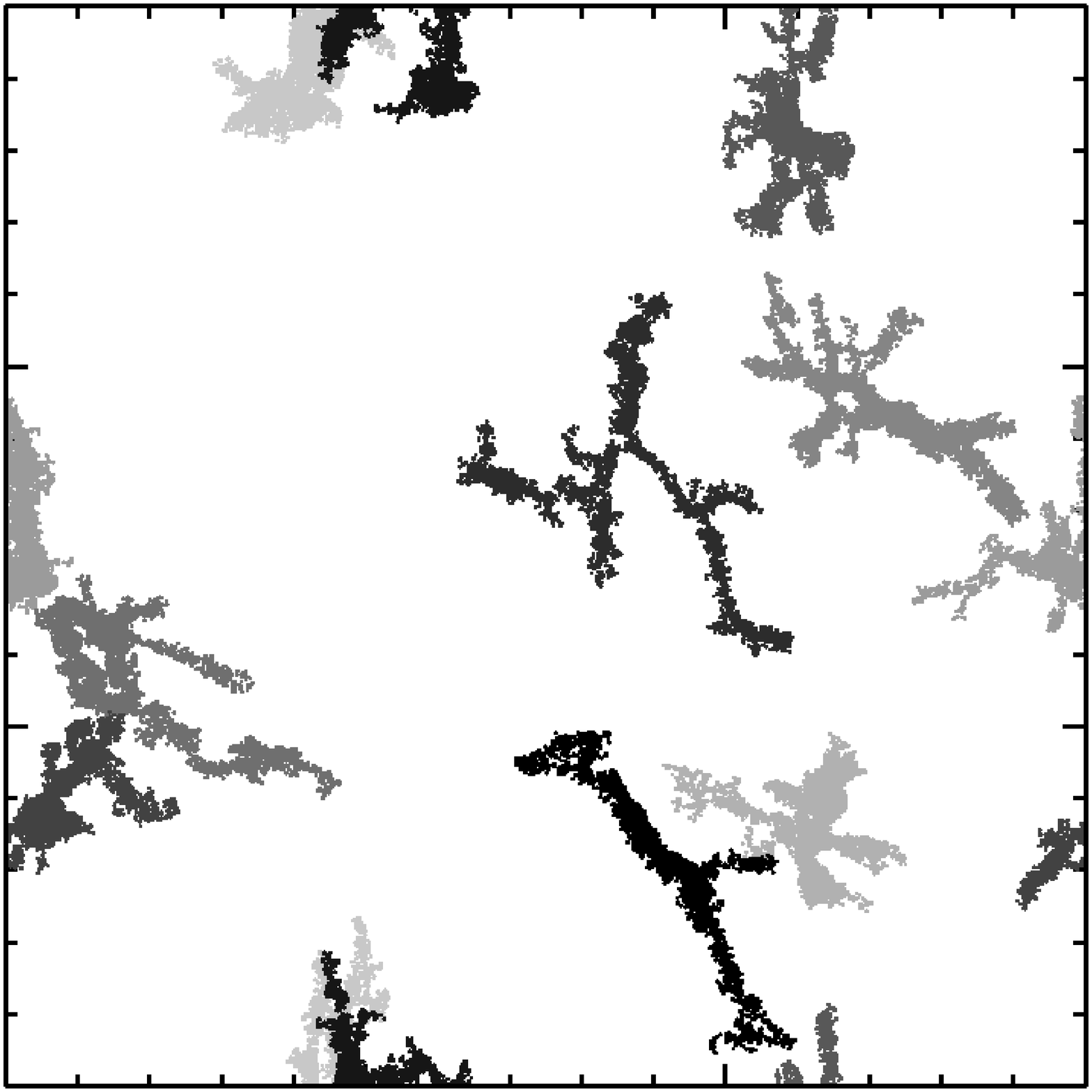}
  \includegraphics[width=0.42\textwidth,angle=0.0]{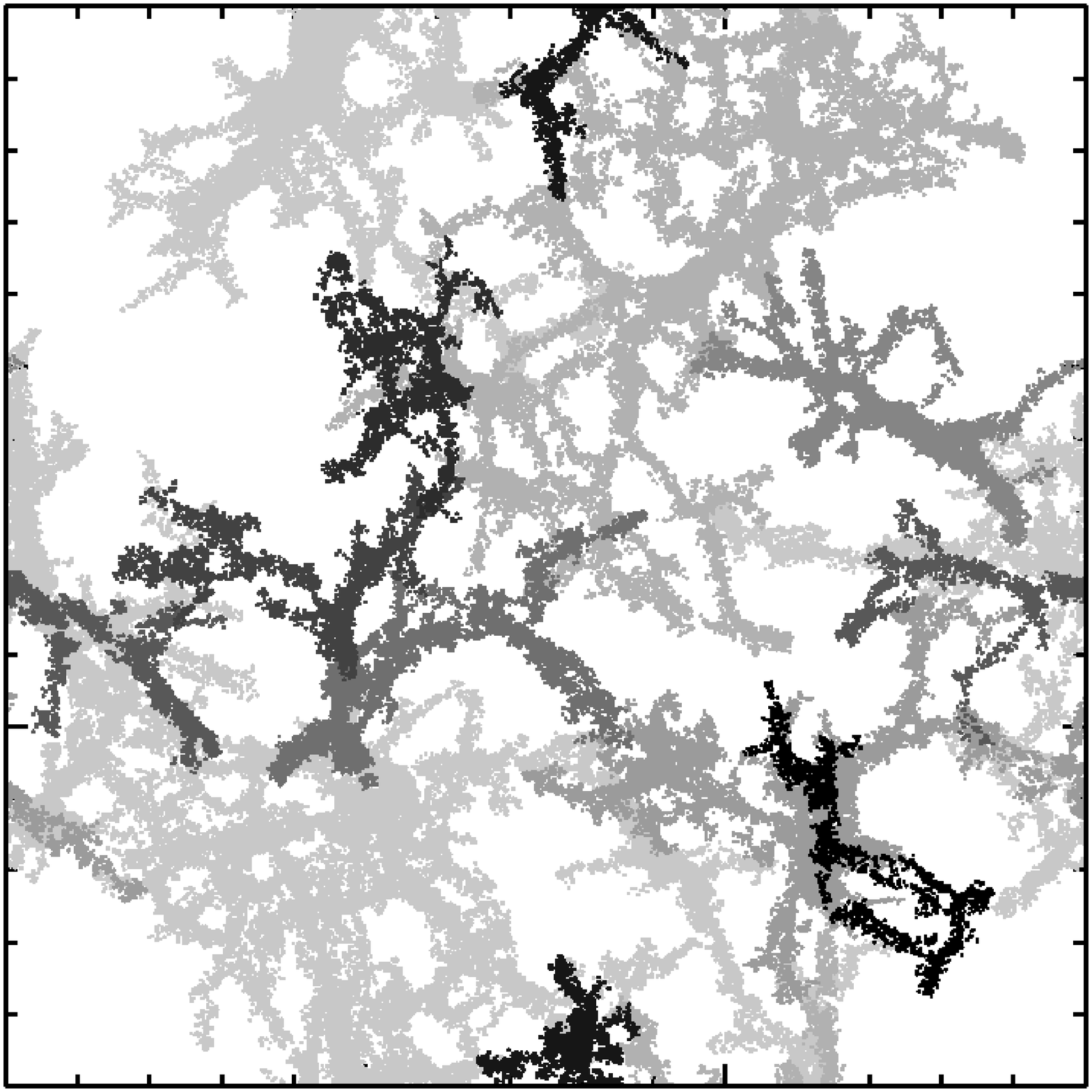}
  \includegraphics[width=0.42\textwidth,angle=0.0]{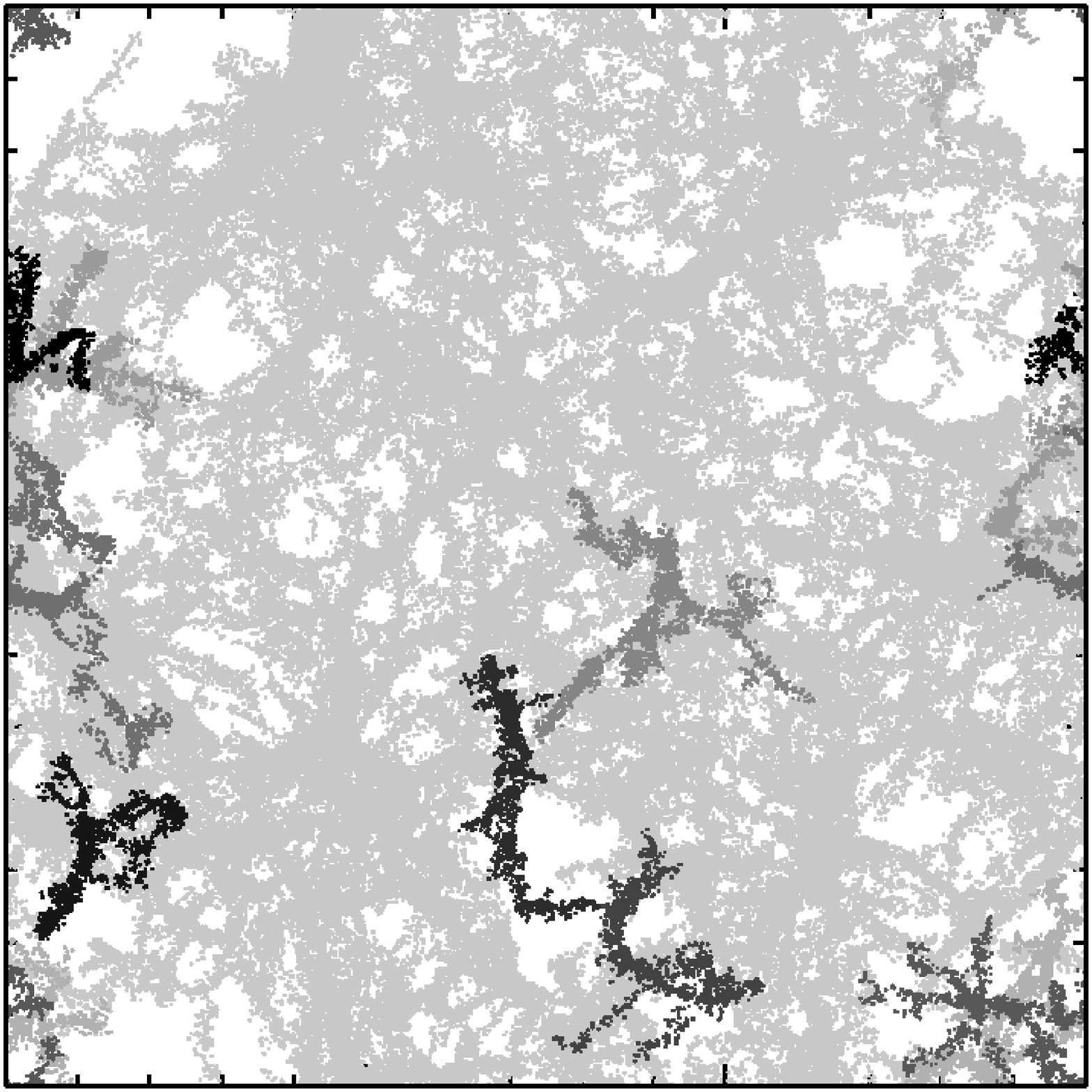}
  \includegraphics[width=0.42\textwidth,angle=0.0]{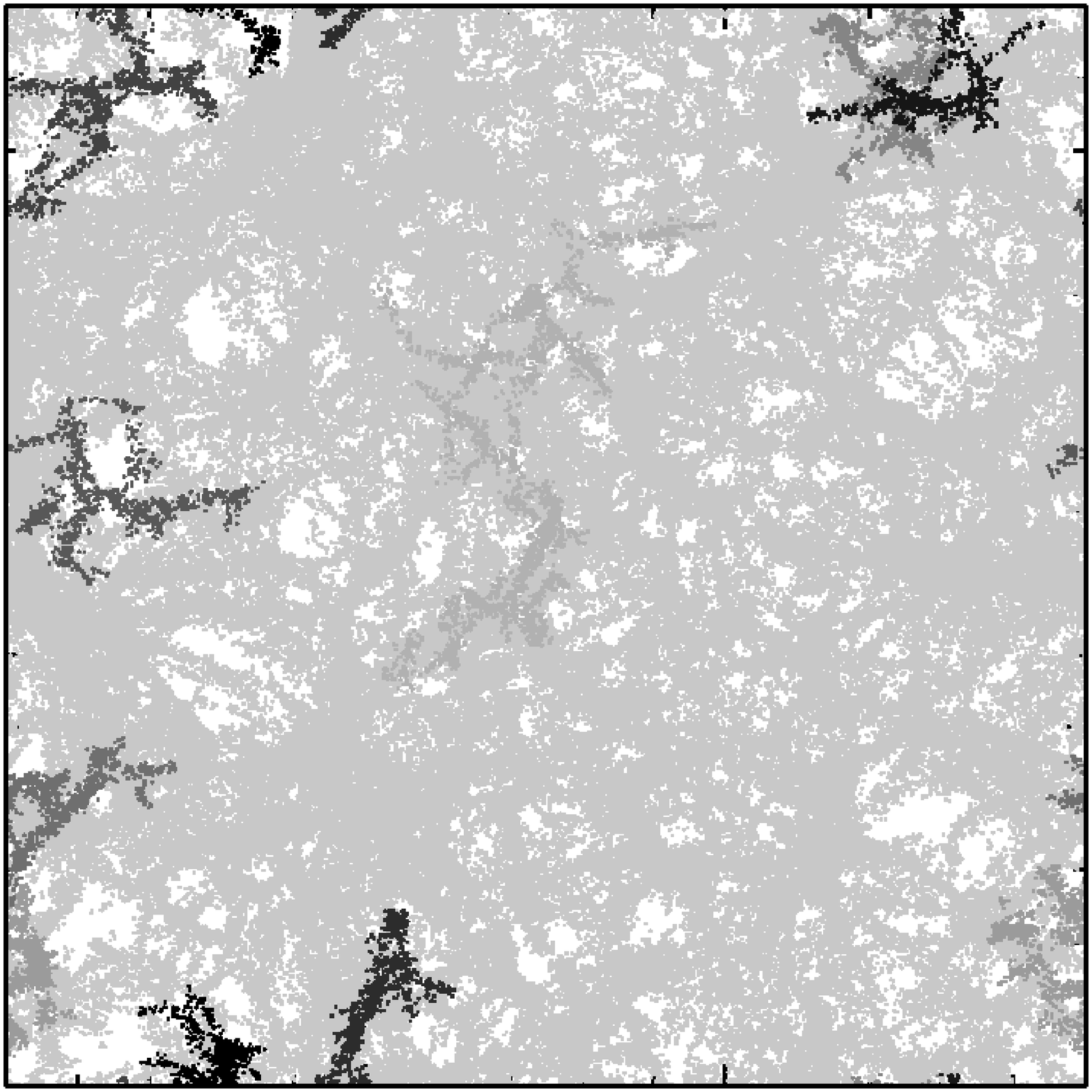}
  \caption{Filament Percolation and Connectivity. The 10 most massive filamentary structures at different density 
           contrast thresholds, $\delta_{th} = 0.2, 0.4, 0.9, 1.8, 2.9, 4.2$ (from top to bottom and left to right).
           In order to differentiate between them, we plot each structure with a different gray tone. 
           The lighter shades correspond to more massive structures. See text for the description.}
  \label{fig:filament_percolation_box}
\end{figure*}
\section{Filaments in the Cosmic web }
\label{sec:filament}
Without doubt, the most salient features of the cosmic web are the 
large filamentary networks, which are interconnected across tens 
and even hundred of Megaparsecs. In this section we focus specifically on 
the filaments identified with the MMF and study their general properties 
such as length, density profile, connectivity, etc. 

\subsection{Filament and wall compression}\label{sec:filament_compression}
Filaments and walls have a complex topology. Their general shape 
may be far from idealized lines and planes. Properties such as direction, density profiles, 
extent and other measures derived from these quantities are difficult to
interpret or meaningless without a proper reference point.
We address this problem by defining their ``heartline" in a similar 
way as the center of mass in spherical clusters is used as reference point.
We  define the one and two-dimensional counterparts for
filaments and walls. They will be referred  as 
the ``spine" of filaments and the ``plane" of walls.

In order to infer the idealized lines and planes from the complex filaments and
walls we performed an iterative algorithm that compresses structures
along their perpendicular direction (normal to the spine of the filament or the plane of the wall)
by moving each particle (or halo) to the center of mass inside a spherical window 
centered on the particle until its position converges. 
The movement of the particles is restricted along the perpendicular 
direction to the spine of the filament or the plane of the wall (see appendix~\ref{sec:compression} 
for details). This procedure \textit{enhances} filaments and walls compressing them closer to
idealized structures: filaments become one-dimensional lines while walls are compressed to nearly 
planar two-dimensional planes (see fig.~\ref{fig:slice_filas_compress_zoom}).

In the determination of spines and planes we based ourselves on dark matter haloes 
instead of particles. Spines and planes derived from the raw
dark matter particles tend to cross the centers of large haloes since they contain most
of the matter in the neighborhood. This gives a similar result as with the use of haloes. 
However, in computing the density profile the difference between particles and haloes can 
become important. The density profile is dominated at small scales by large haloes, giving the 
false impression of highly dense cuspy cores or even worst, producing several ``cores'' of
a single filament \citep{Colberg05}.

\subsection{The filamentary network}\label{sec:filamentary_network}
Figure \ref{fig:filamentary_network} shows a slice of the simulation box in which 
filaments have been compressed to delineate their spines. Gray circles
indicate the location of clusters with masses above $10^{14}$ M$_{\odot}$ h$^{-1}$. This figure
presents the Cosmic Web as a network of interconnected filaments spread all over the
simulation box. The clusters sit at the intersections or ``nodes" of the network.
The filamentary network permeates all regions of space, even the very 
underdense voids. One important aspect of the filamentary network is its
cellular nature \citep{Joeveer78,Zeldovich82} which defines a multiscale system marked
by structures over a range of scales \citep{Sheth04,Shethwey04,Shen06}. 
Large voids are delineated by thick large filaments. Each of these voids contain subsystems of smaller 
filaments delineating smaller mini voids which in term form the basis for even smaller systems 
(\citet{Dubinski93,Schmidt01,Shethwey04,Shen06,Aragon10}). Also, large empty voids contain extremely 
tenuous but rich filamentary systems only seen in high resolution simulations 
\citep{Weykamp93,Gottlober03,Colberg05b}. 

\subsection{Filamentary Network:\\ \ \ \ \ \ \ \ Percolation and Connectivity}
\label{sec:filpercolate}
The raw output of the MMF is an Object Map $\mathcal{O}$, which defines which pixels 
belong to a given morphology \citep[see][]{Aragon07b}. Given the coherence of the filamentary 
(and sheetlike) patterns in the Cosmic Web, such pixels connect with each other into large 
volume pervading complexes. This network connects filaments of a large variety of 
sizes, in turn branching into smaller filaments which are often multiply connected. 

Such connectivity information, as well as the individual properties of filaments, is not explicit 
in the Object Map. To this end, we assess the percolation behaviour of the \textit{density field contained 
in the filamentary network}. This avoids the ambiguity in the segmentation of morphologies based 
on density alone (see sect.~\ref{sec:densseg}). 

The percolation analysis studies the change in number, properties and/or connectivity of objects defined as 
regions of space above a given density contrast threshold $\delta_{th} \equiv \rho_{th}/\bar{\rho}-1$. By 
varying the threshold across the complete range of density values in the matter distribution, we obtain 
a systematically evolving population of structures, each characteristic for the value $\delta_{th}$ 
\citep{Zeldovich82,Shandarin83a,Shandarin83b,Klypin88}. 

\begin{figure}
  \centering
  \vskip -0.5truecm
  \mbox{\hskip -0.5truecm\includegraphics[width=9.0truecm]{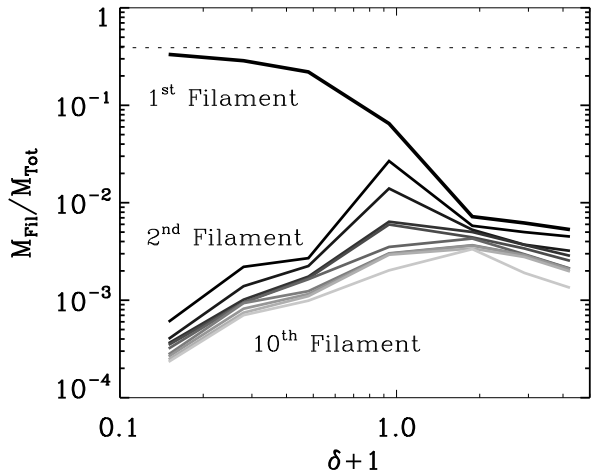}}
  \vskip -0.5truecm 
 \caption{Mass ratio between the $1^{th}$, $2^{nd}$ up to the $10^{th}$ largest filaments and the total mass in the
           simulation box. The horizontal dotted line indicates the total mass content in 
	   filaments (see table~3).}
  \label{fig:filament_most massive_statistic}
\end{figure}

We assess the change in filamentary complexes as a function of the density contrast 
threshold:
\begin{equation}
    1+\delta_{th}\,\equiv\,\left( \frac{4}{3}\pi l_{link}^3 \right)^{-1},
\end{equation}
\noindent where $l_{link}$ is a linking length between two dark matter particles. By iteratively associating 
particles with separations $d \le l_{link}$ we produce a catalogue of filamentary complexes. This procedure 
is rather similar to the identification of clusters using the friends of friends algorithm. Following their 
identification, we rank the filamentary configurations by their mass, ie. the number of particles they contain. 
This results in a mass ordered list consisting of the most massive filament, second most massive filament, 
third most massive, etc. 

At high densities, filaments are isolated objects with a simple shape and topology. As the value of $\delta_{th}$ 
decreases, the filaments grow steadily while more mass from their surroundings is added to them\footnote{By restricting ourselves 
to the filament pixels, the structures remain confined to filaments and do not flood into walls or voids.}. While they 
grow they branch into increasingly complex structures. At some point, at the merging threshold 
$\delta_{th} \sim 2$, there is an rather sudden transition in the way the filaments grow. No longer the steady inclusion 
of mass from adjacent lower density regions constitutes the main growth process. Instead, the merging of existing filamentary 
complexes into super-filaments becomes the main mode of structure growth. Descending to even 
lower densities, the filaments continue to merge until, rather abruptly, at one particular 
density value a single superstructure emerges which spans the entire volume: this marks the 
{\it percolation} transition. As a result, opposite faces of the simulation box are connected. 

The growth process is illustrated in figure~\ref{fig:filament_percolation_box}. It follows the 
development of the 10 most massive filaments along a range of decreasing density thresholds 
$\delta > \delta_{th}$ (going from top left to bottom right). The figure shows the filaments 
at thresholds $1+\delta_{th}= 0.2, 0.4, 0.9, 1.8, 2.9$ and $4.2$ (from top to bottom and left 
to right). In order to distinguish them, each of the filaments is plotted with a different 
gray tone, with the lighter shades corresponding to more massive structures. The panels highlight the 
non-linear nature of the percolation process, with the initial gradual growth suddenly transiting into 
the merging of filaments and the emergence of super complexes. 

The corresponding growth in mass of the 10 most massive filaments, as a function of 
threshold density $\delta_{th}$, is plotted in figure~\ref{fig:filament_most massive_statistic}. 
It depicts the mass fraction of each of these filaments. At high values of $\delta_{th}$, the 
10 filaments have similar masses, confirming the impression obtained from 
fig.~\ref{fig:filament_percolation_box}. At low, percolating, density values there 
is a distinct difference between the largest filament and all other filaments. Towards 
the lowest density values, $\delta \approx -0.8$, the most massive filament has absorbed 
the major share of filamentary objects. It asymptotically attains a mass of $\sim 40\%$ of 
the total mass, which is the entire mass enclosed by the filamentary network (see 
table~3). Meanwhile, the mass of the remaining filaments 
decreases continuously. As their more massive peers get absorbed into the percolating 
principal filament, the remaining isolated objects represent ever smaller specimens 
of the filament population. 

The largest structure in the percolation process, referred to as the \textit{principal percolating filament},  
\footnote{We use the term filament since it is contained inside the filament Object Map \citep[see][]{Aragon07b}}. 
carries important information on the topology of the density field \citep{Shandarin04}. The principal percolating 
filament has significantly different properties than the rest of the (much smaller) filaments. It is a space-covering 
network connecting all regions of space and hardly changes significantly once it has reached the 
percolation threshold. 
 
\begin{figure*}
  \centering
  \vspace{1.5cm}
  \mbox{\hskip 0.4truecm\includegraphics[height=9.0cm]{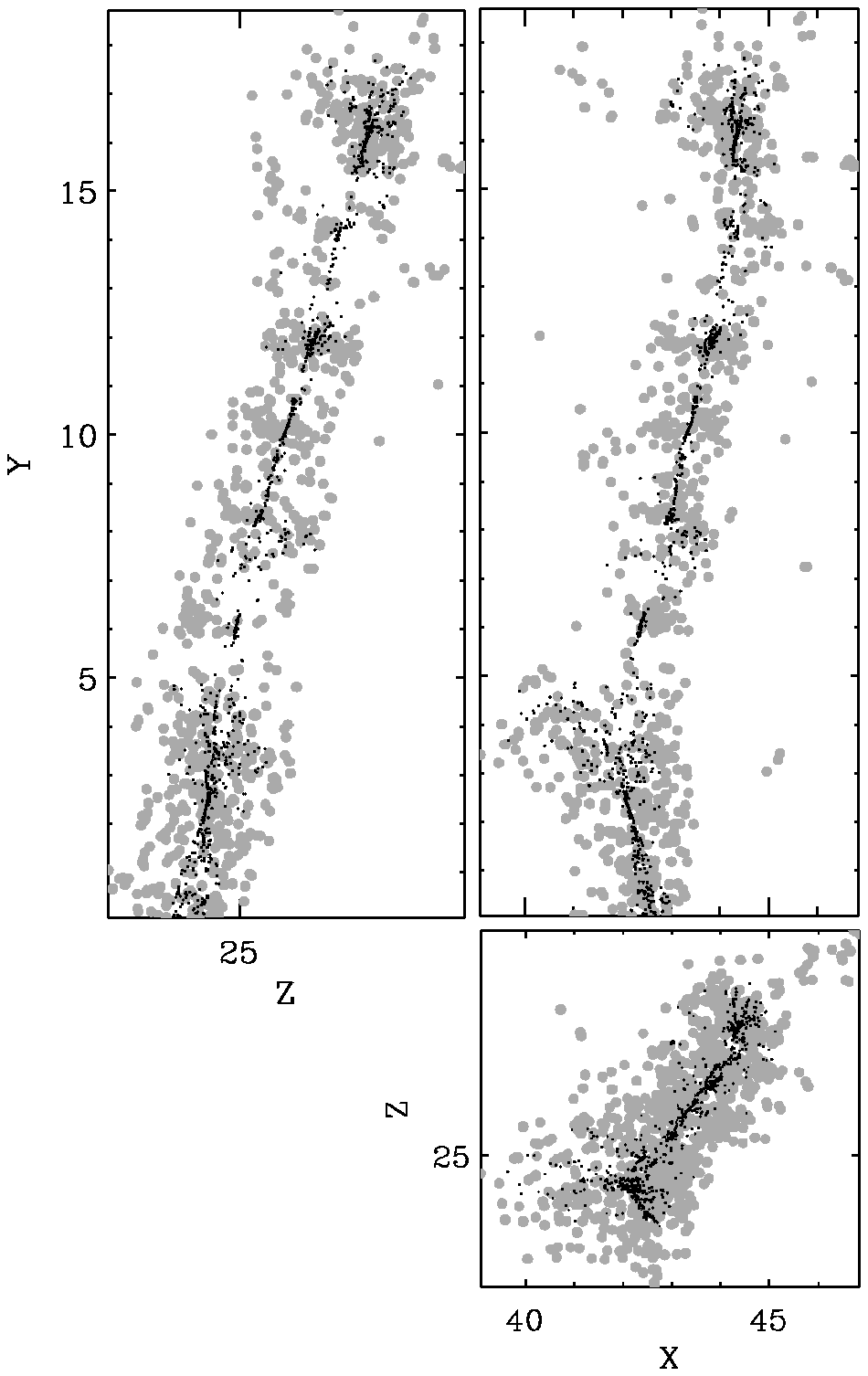}}
  \mbox{\hskip 2.5truecm\includegraphics[height=9.0cm]{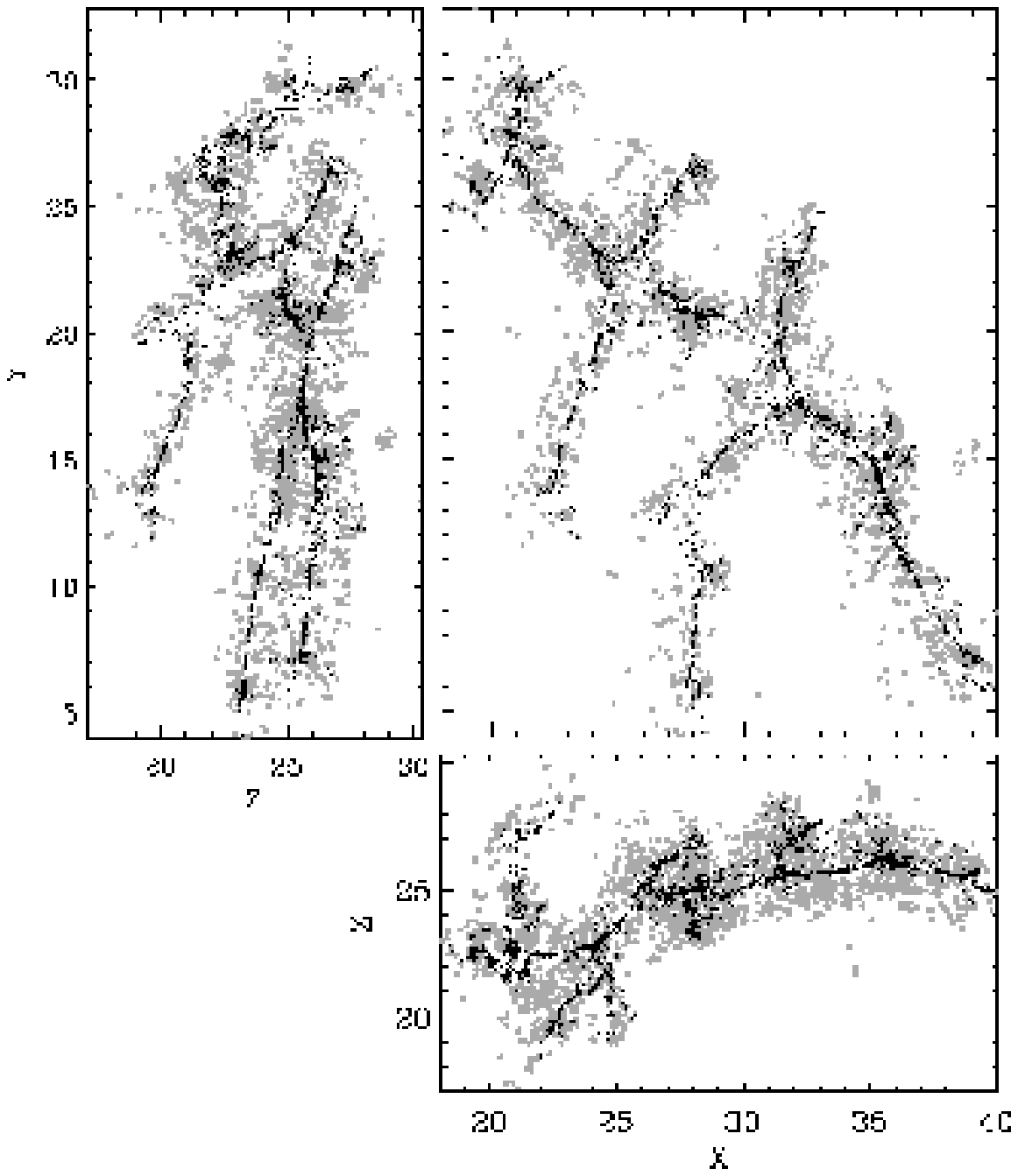}}
    \caption{Filament Diversity. Top: three orthogonal projections of typical \textbf{line} (top) and \textbf{star} (bottom) filaments. 
             Dark matter particles are indicated by filled gray circles. The spine of the
	     filament is also shown, by black dots, in order to better delineate the shape of the
	     filaments. Bottom: three orthogonal projections of typical \textbf{grid} (top) and \textbf{complex} (bottom) filaments. 
             Dark matter particles are indicated by filled gray circles. The spine of the
	     filament is also shown, by black dots, in order to better delineate the shape of the
	     filaments.}
  \vspace{0.5cm}
  \mbox{\hskip -0.7truecm\includegraphics[height=8.0cm]{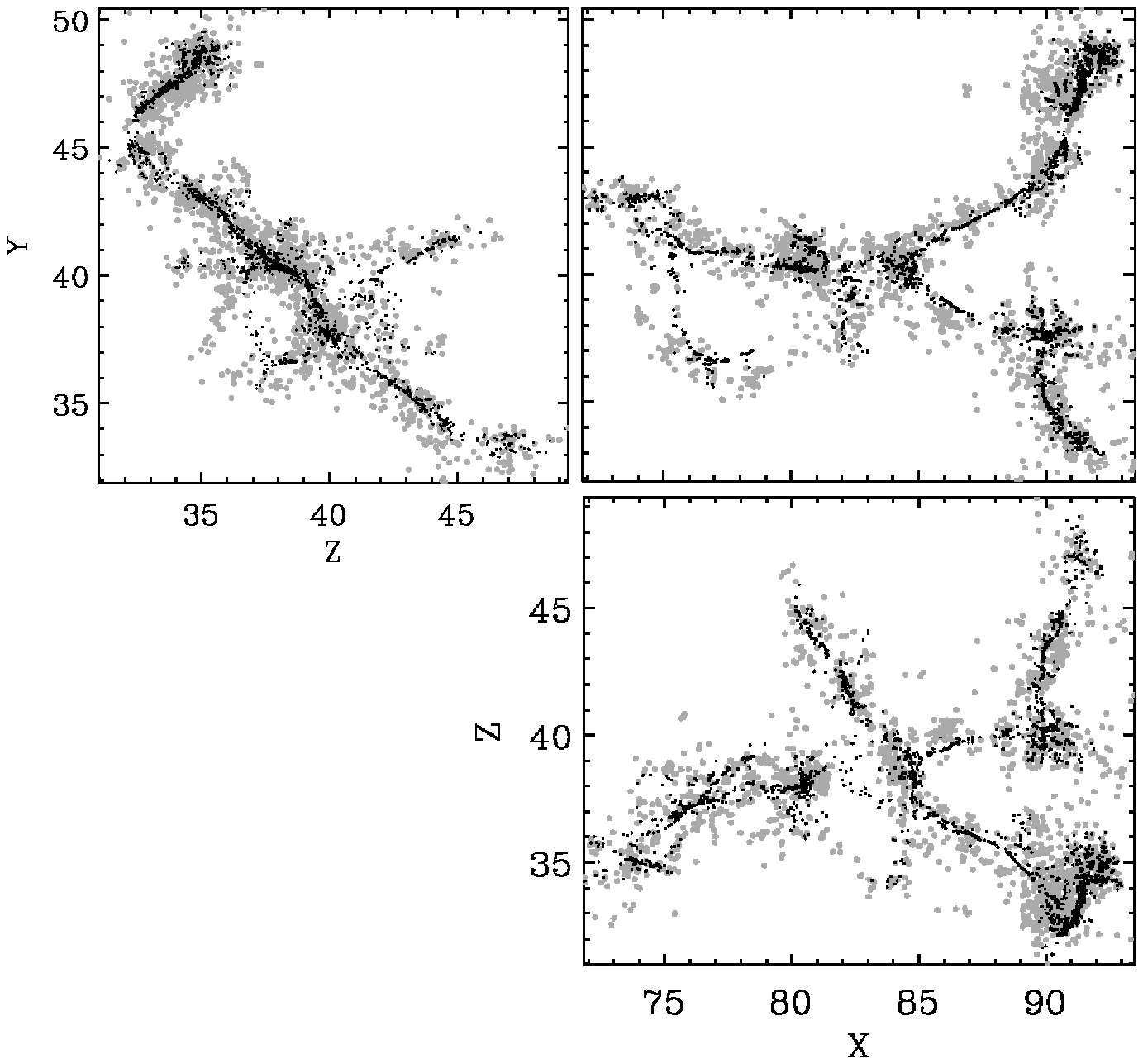}}
  \mbox{\hskip 0.7truecm\includegraphics[height=8.0cm]{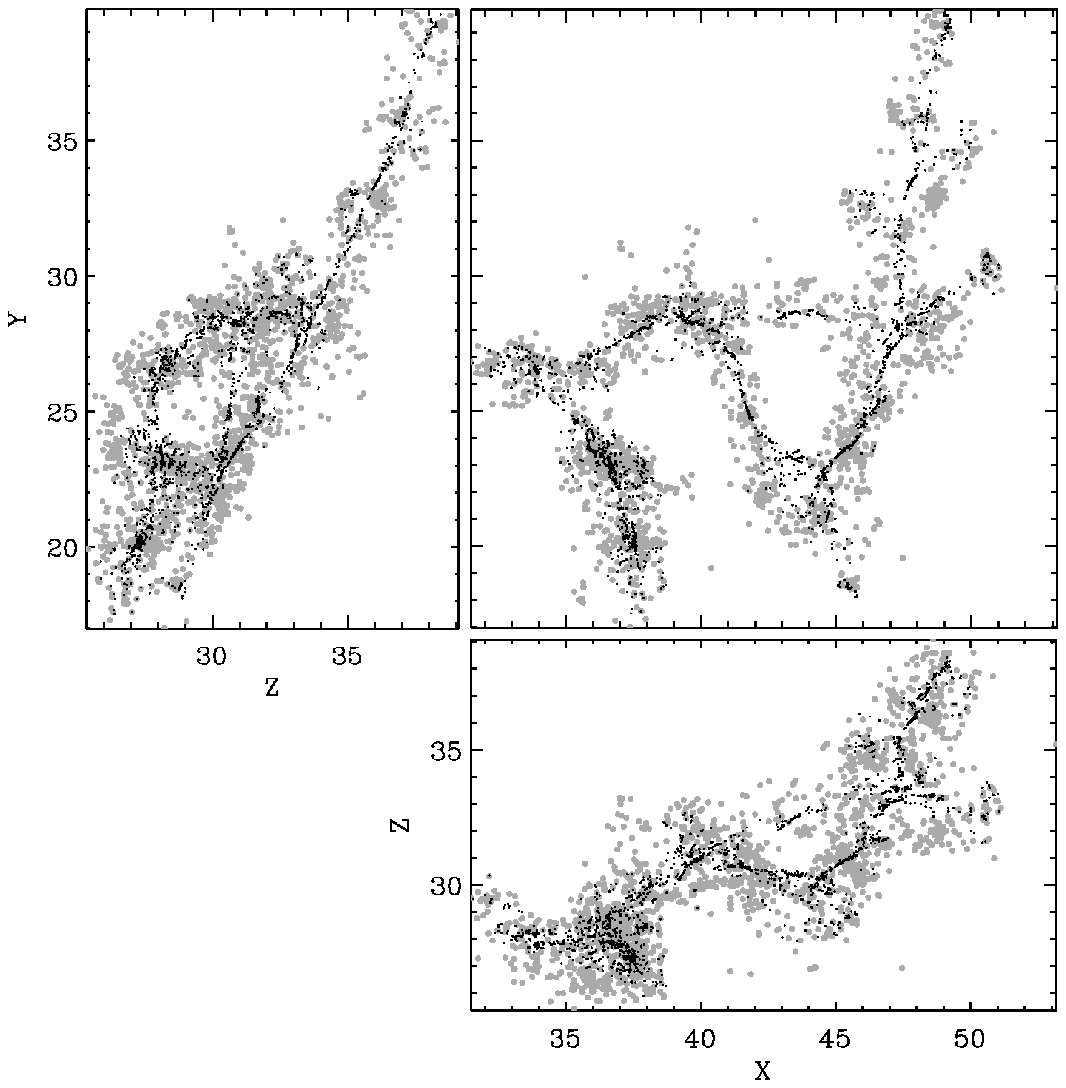}}
  \vspace{1.5truecm}
  \label{fig:filaments_compressed_fif1}
\end{figure*}
\subsection{Isolating Individual Filaments}
\label{sec:filid}
The ability to recognize individual features such as filaments is natural to 
the human brain. However, the analogue for computational recognition still represents a major challenge. 
To \textit{identify and isolate} the individual elements forming the interconnected network, 
we need to invoke post-processing procedures. This involves the definition and introduction of 
user-specified measures. 

As we have seen in previous sections, strictly speaking the filamentary network is a system that connects 
\textit{all} filamentary features. In this sense it does not constitute a sample of individual isolated 
structures, so that any attempt towards the identification of individual filaments necessarily involves a 
level of subjectivity. And even though the MMF formalism provides us with an objective measure of 
filamentariness at each position of space, it remains far from trivial for MMF to dissect the filamentary 
network into objectively defined individual filaments. 

One strategy to dissect the filamentary network into individual objects is by exploiting its percolation 
properties. Following the argument of \citet{Shandarin04} that individual objects identified by means 
of density thresholds should be studied \textit{before percolation occurs}, we use the same principle 
to select the density threshold for defining individual filaments. 

From figure~\ref{fig:filament_most massive_statistic} we see that the merging density threshold, 
below which the identified features start to merge with each other, is in the order $\delta_{th} \sim 1$. In practice, we 
use a somewhat larger value of the threshold in order to eliminate the possibility of filament 
mergers via thin tenuous bridges whose significance might be questioned (this is a known problem of 
the friends of friends algorithm for halo detection \citep{Eisenstein98}). Visual inspection indicates 
that the differences are small, a consequence of the restriction to regions identified by MMF as 
filaments. After some experimentation, we use the value $\delta_{th} = 3$ as the density threshold 
for the identification of individual filaments.

\begin{figure}
  \centering
  \mbox{\hskip -0.5truecm\includegraphics[width=0.5\textwidth,angle=0.0]{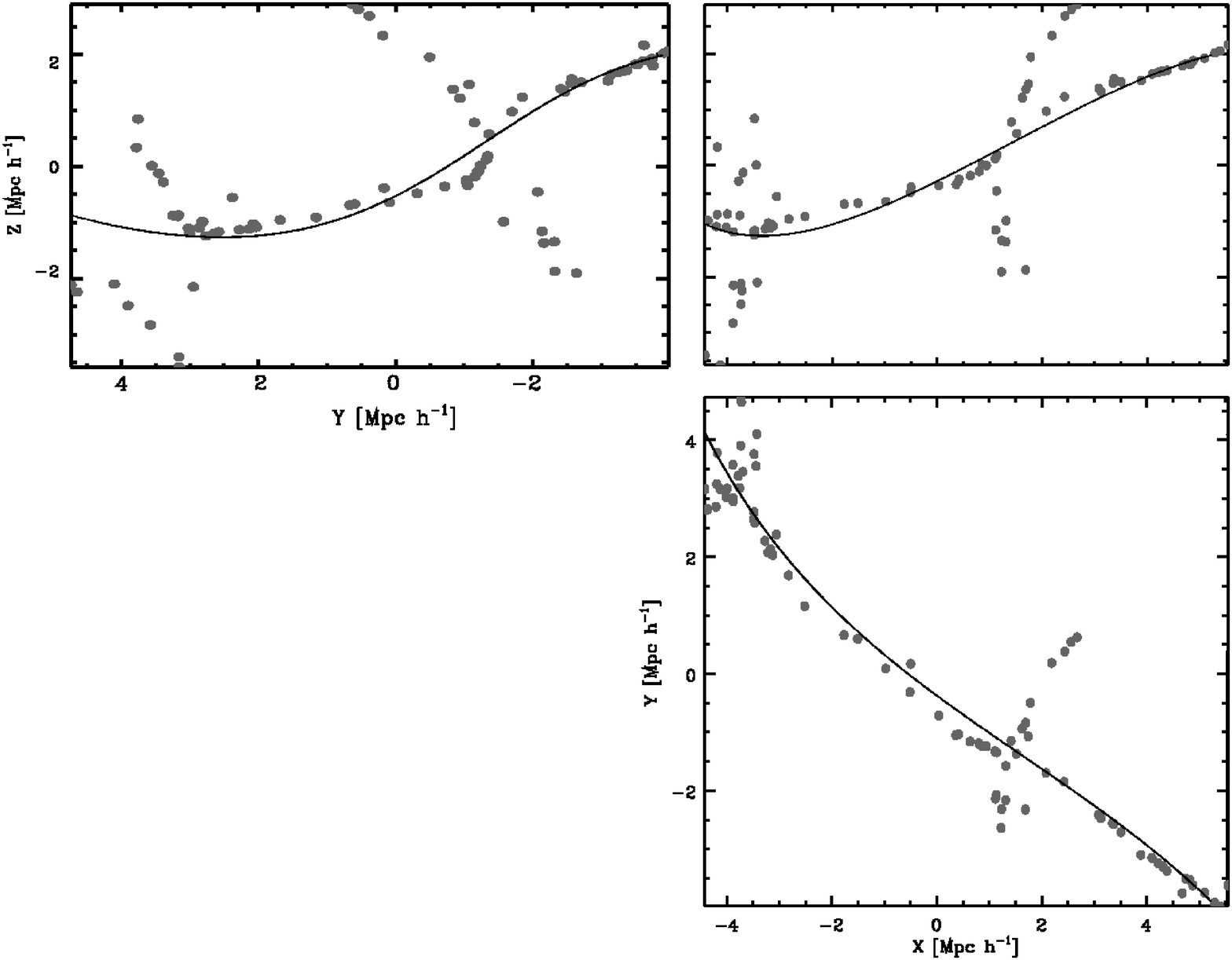}}
  \vskip 0.75truecm
  \mbox{\hskip -0.0truecm\includegraphics[width=0.5\textwidth,angle=0.0]{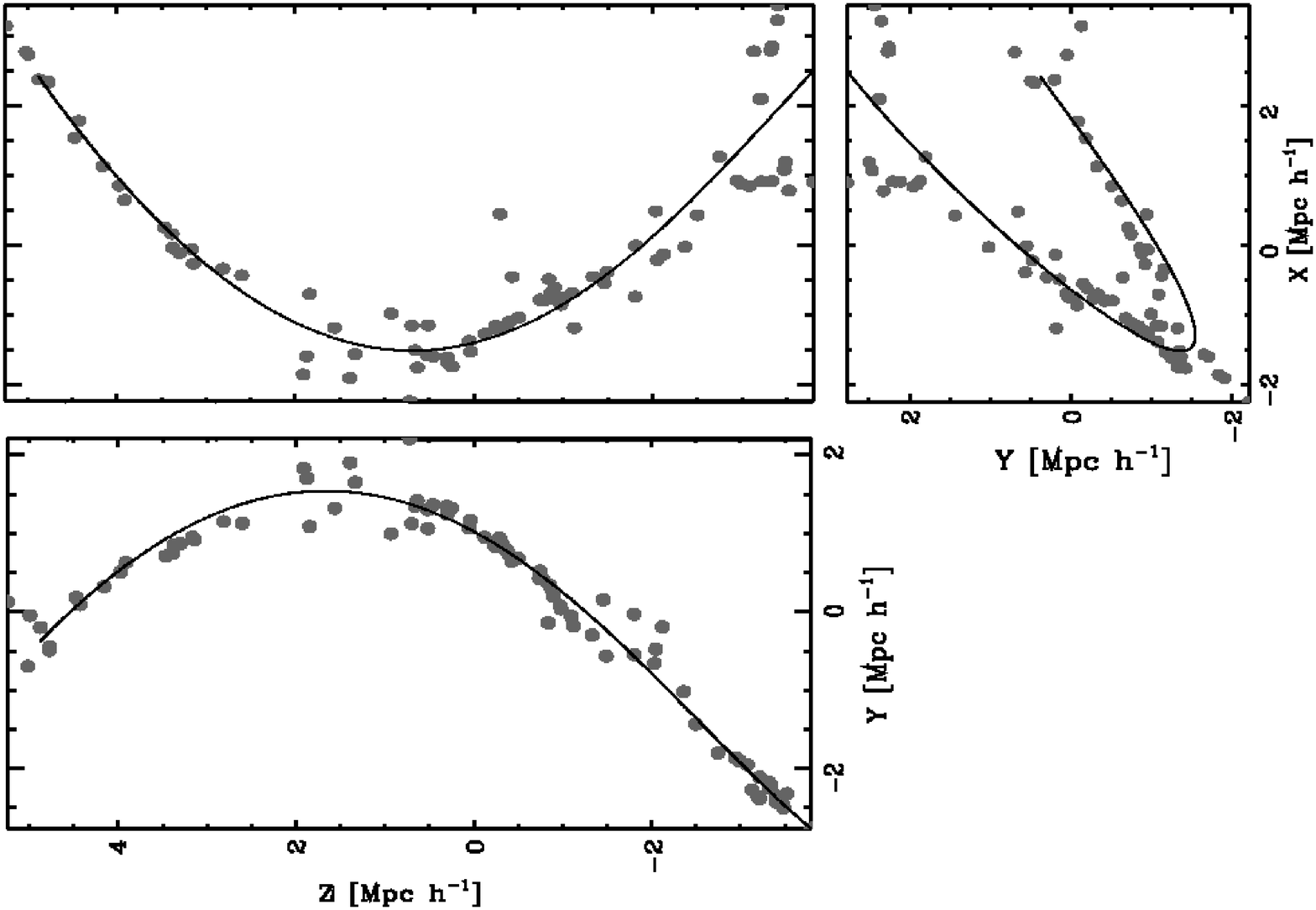}}
    \caption{Polynomial fit (solid line) of two of the largest filaments in the simulation. 
             Dark matter haloes are represented by gray circles.}
  \label{fig:filament_polynomial_fit}
\end{figure}
\subsection{Filament classification}
\label{sec:filament_types}
Figures~\ref{fig:filaments_compressed_fif1} show four
examples of typical filaments. For visualization purposes we plot dark matter particles taken 
from the $128^3$ simulation in grey color. Superimposed on these we plot the spine of the filament 
by means of the black particles. Haloes delineate the same structures, be it more diluted.
In general we find that the level of complexity of filaments is related to the surrounding large 
scale matter configuration. Filaments in the vicinity of massive clusters form more complex systems 
than the ones connected to less massive clusters.

The fractal nature of the filamentary network makes it difficult to classify individual filaments 
since in principle they form a percolating network that includes all filaments. Also, the branching properties 
of our filaments ultimately depends on the resolution limit of our simulation. This is a natural consequence of 
the hierarchical development the Cosmic Web \citep{Sheth04,Shethwey04,Shen06}. Ultra-high resolution N-body 
simulations show that even in the most underdense regions one can find systems of tenuous filaments extending 
along the whole physical extent of the voids \citep{Weykamp93,Gottlober03,Platen07}. 

On the basis of a rough phenomenological inventory of the shape and morphology of the filaments in our simulations, 
we distinguish four basic types of filaments:

\begin{enumerate}
\item[$\bullet$] \textbf{Line filaments} do not have branches (or very few) and are mostly 
      straight with lengths in the order of 5-20$\Mpch$. They are often 
      found as ``bridges" between massive clusters. Shorter filaments are also
      straighter than large ones. They may be identified with the 
      \textit{intracluster filaments} studied by \citet{Colberg05} and 
      filaments type 0,I,and II in the classification of \citet{Pimbblet04b}.\\
\item[$\bullet$] \textbf{Grid filaments} are often found crossing vast regions with no
      massive clusters crossing them. They form the surrounding ``net" enclosing large voids
      and are almost invariably two dimensional, suggesting that walls are in fact
      delineated by these kind of filaments.
      Even when they consist of several branches one can often identify a main ``path" with 
      smaller filaments running almost perpendicular to it.\\
\item[$\bullet$] \textbf{Star filaments} have a well defined ``center", usually a cluster
      or large group from which several ``arms" stretch out.
      Star filaments can be considered a smaller version of grid filaments.
      They are also two dimensional structures, suggesting that grid and star filaments
      represent the same kind of structures.\\
\item[$\bullet$] \textbf{Complex filaments} do not have a clear shape, they are often
      multiply connected and it is difficult to define a main path or
      direction. These filaments can be found in almost any environment.
\end{enumerate}

\subsection{Length of filaments}
\label{sec:length_of_filaments}
While describing filaments in terms of their mass is straightforward 
(see section \ref{sec:filpercolate}), the determination of their length 
and related physical properties involves several complications because 
of the branching nature of the filamentary network. The length of 
complex systems composed of several interconnected branches is not 
straightforwardly or uniquely defined. One may even argue that it is a
rather meaningless concept in the case of filaments with strong branching 
such as grid and star filaments, for which it is almost impossible to 
define a main spine. 

\begin{figure}
  \centering
  \vskip -0.6truecm
  \includegraphics[width=0.48\textwidth,angle=0.0]{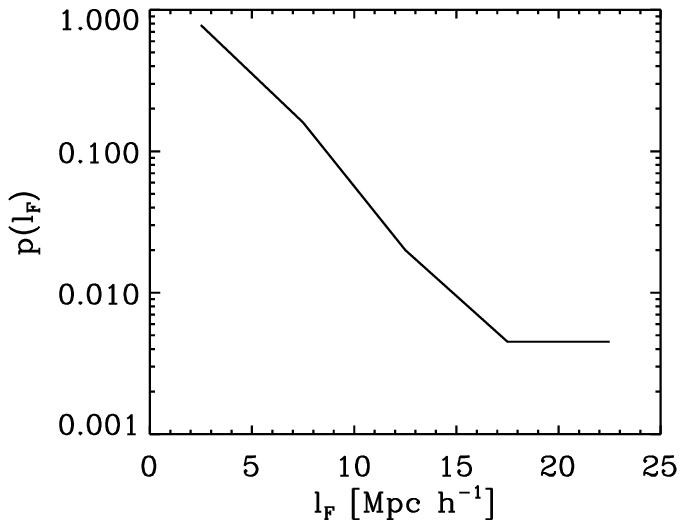}
  \vskip -0.5truecm
  \includegraphics[width=0.48\textwidth,angle=0.0]{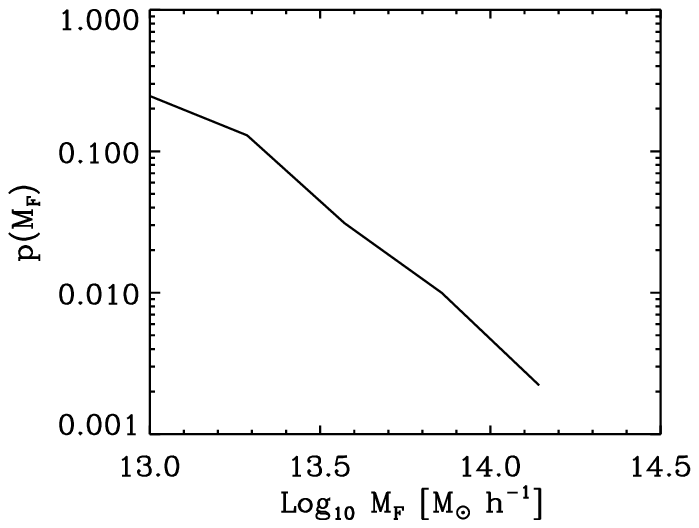}
    \caption{Probability density distribution of length (top panel) and mass (bottom panel) of filaments
             (see text for details). Only filaments with more than 10 haloes were considered.}
  \label{fig:filas_size_distribution}
\end{figure}
\bigskip
We may consider various options for the definition of the length of a filament:
\begin{itemize}
\item[$\bullet$] The total length of all the branches of filament, related to the fractal nature 
of the filamentary network \citep{Martinez90}. Its definition presents several practical 
complications such as the identification of the branching points in the main path.
\item[$\bullet$] The length of the main path of the filament \citep{Colberg05,Pimbblet04b}. 
This definition presumes the existence of such a main path, whose definition introduces 
an arbitrary choice of defining criteria. 
\end{itemize}
\noindent Given the high level of complexity of the algorithms used to identify the total
length of filamentary systems we postpone their study for future work. Instead, here we 
concentrate on the length of the main path of the filament. To first order this can be 
identified with the thickest or longest branch.

We are particularly interested in the length of filaments connecting 
to clusters. These do not only appear as the nodes of the Cosmic Web, but they 
also define the formation sites of filaments \citep{Bondweb96}. They 
therefore provide a natural way of dissecting the filamentary network. 
Accordingly, we proceed different than in the percolation-based dissection 
described in sect.~\ref{sec:filid}. For the construction of the 
filament catalogue, we use the haloes instead of the dark matter particles to 
trace the filamentary network. This has the advantage of being faster when 
computing the length of the filaments and of ignoring the irrelevant small 
details. 

\begin{figure*}
\mbox{\hskip -0.5truecm\includegraphics[width=0.99\textwidth,angle=0.0]{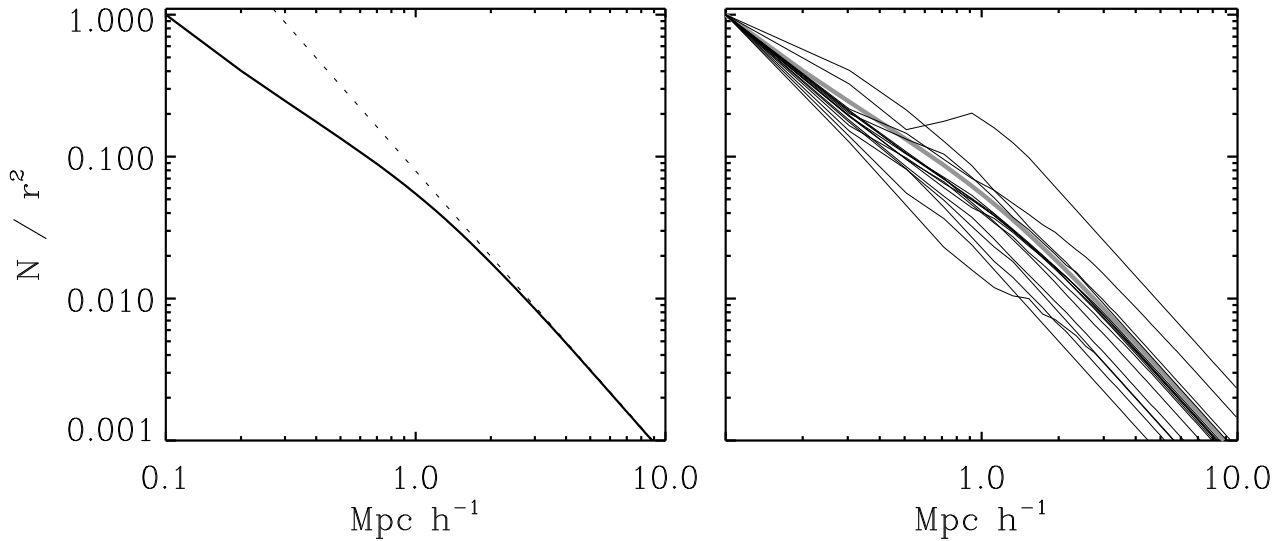}}
\caption{Filament Density Profile. Left: average enclosed density profile of haloes inside a filament. The dotted line 
         corresponds to a power-law fit $N(r) \propto r^{-2}$. Right: density profiles of 15 individual filaments. 
         Superimposed (thick gray line) is the average enclosed density profile. To enable their mutual comparison, all 
         density profiles are scaled to a value $N(r=0)=1$ at the center of the filament.}
\label{fig:filaments_density_profile}
\end{figure*}
We start by collecting all clusters with a mass $M \ge 10^{14}$M$_{\odot}$ h$^{-1}$. 
Around the locations of these clusters, we cut spheres with radius 2 $\Mpch$. The 
filaments contained in these spheres involve smaller individual objects, which are 
easier to handle. From this set of isolated filaments we produce a catalogue using 
a friends of friends algorithm with a linking length corresponding to a density 
threshold of $\delta_{th}+1= 4$.  Note that in this way we exclude large 
superstructures, as the filaments are broken up at cluster nodes.

To analyze the length of the identified filaments, we use a 3$^{rd}$ order polynomial fit to 
describe and quantify their shape (see appendix.~\ref{sec:fillength} for details). Two representative 
filaments are shown in fig.~\ref{fig:filament_polynomial_fit}, along three mutually orthogonal 
directions. The best polynomial fit is superimposed on the related halo distribution. The fit manages 
to closely follow even the most intricate filaments and also ignores small branches. This is particularly 
visible in the case of the \textit{grid} filament in the top panel, whose main branch is crossed 
by several smaller filaments. 

The length and mass distribution of the resulting filament sample is shown in 
fig.~\ref{fig:filas_size_distribution}. Small filaments are clearly more abundant 
than the large ones, as we may expect for a hierarchically evolved distribution. Filaments with 
lengths in excess of a few tens of Megaparsecs are extremely rare. In terms of their mass, 
we also see that there are hardly any filaments with masses larger than $\sim 10^{14}$ M$_{\odot}$ h$^{-1}$. 
In other words, while the largest and most massive filaments are the most prominent 
structures in the Cosmic Web, they represent only a minor fraction of the entire 
filament population. 

\subsection{The density profiles of filaments}
\label{sec:density_profiles}
Filaments are far from being smooth uniform structures. In most cases
it is possible to identify a highly dense spine surrounded by more
diffuse matter. Filaments are also populated by compact dense haloes.  
This yields a resemblance of filaments to a pearl necklace, with 
haloes suspended along the bridging spine between massive clusters. 
Despite the prominence of the inner realms of filaments, there are only 
a handful of studies adressing their density profile with respecty to the 
filament's spine \citep[e.g.][]{Colberg05,Dolag06}. This is even 
more true for the density distribution of walls, which has only 
been adressed in a few theoretical studies \citep[see e.g.][]{Zeldovich70,Shandzeld89}.

\begin{figure}
  \centering
  \includegraphics[width=0.4\textwidth,angle=0.0]{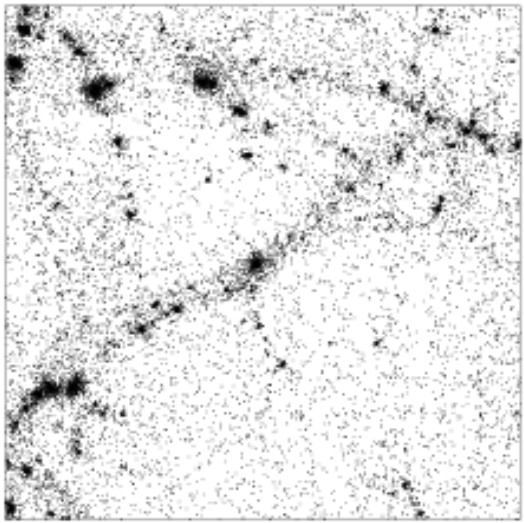}
  \vskip 0.25truecm
  \includegraphics[width=0.4\textwidth,angle=0.0]{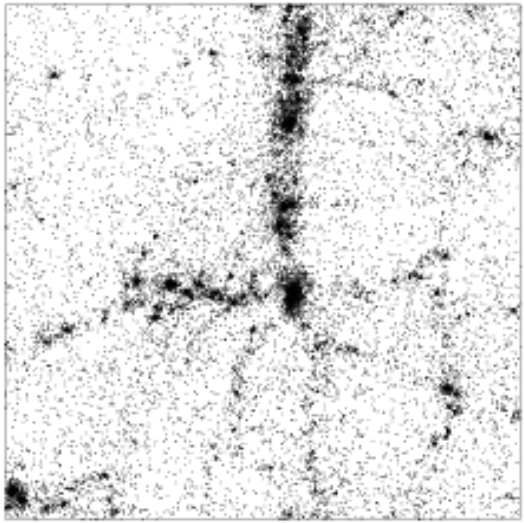}
  \vskip 0.25truecm
  \includegraphics[width=0.4\textwidth,angle=0.0]{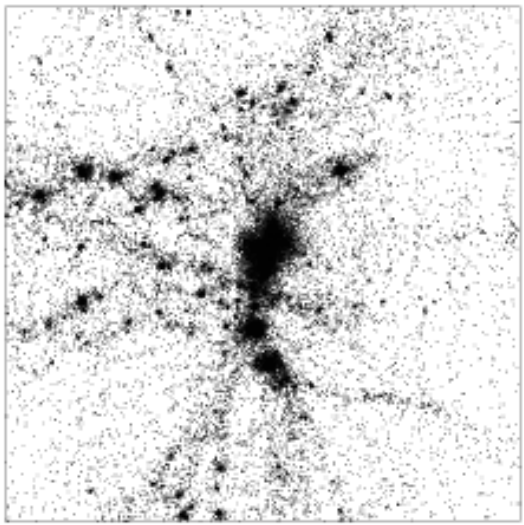}
    \caption{Three examples of clusters and the filaments to which they are connected. The three 
     clusters have masses in the range  $10^{13} - 10^{14}$ M$\odot$ h$^{-1}$. Top: a grouplike 
     cluster, with $M \sim 10^{13}$  M$\odot$ h$^{-1}$, connected to two filaments. Centre: A 
     medium sized cluster, with $M \sim 2-3 \times 10^{13}$  M$\odot$ h$^{-1}$, connected 
     to several filaments. Bottom: a massive cluster, with $M \sim \times 10^{14}$  M$\odot$ h$^{-1}$, 
     at the centre of a highly complex environment of filamentary branches. }
  \label{fig:cluster_filament_connections_particles}
\end{figure}
Visual inspection of the observed Cosmic Web reveals that there are indeed filaments spanning 
across several tens of Megaparsecs. Examples of these are the ``spine'' of the 
Pisces-Perseus supercluster \citep{Gregory81,Giovanelli85} and the planar filamentary 
system known as the Sloan Great Wall \citep{Gott05,Platen09}. However, such massive 
systems are marked by a substantial degree of substructure, containing numerous 
clusters and a filigree of small-scale filaments. 

In general, large haloes dominate the density profile. This manifests itself in a cuspy 
profile centered at the spine of the host filaments, and even occasionally involving several 
peaks near the centre. Following this observation, we use haloes instead of dark matter 
particles, thereby avoiding contamination of the radial profile at small scales. The density profiles of the 
filaments are determined by counting the \textit{enclosed number of haloes}, weighted by their mass, 
in bins of increasing radial distance from the spine of the filament. The radial distance of a halo 
with respect to the spine of its host filament is defined as the displacement of a halo \textit{before} 
and \textit{after} we apply the compression algorithm described in section \ref{sec:filament_compression} 
(see appendix~\ref{sec:compression} for details). To enable the mutual comparison of the density profiles of all 
filaments, they are all scaled to a value $N(r=0)=1$ at the center of each filament. The resulting 
density profile, averaged over all filaments, is shown in the left panel of 
fig.~\ref{fig:filaments_density_profile}. A similar analysis for walls reveals that they have 
less well defined boundaries with widths ranging between 5-8 $\Mpch$. 

The average filament profile has a power-law shape, $N(r)\propto r^{-2}$, beyond a radius of $r\sim 2 \Mpch$. 
For a one-dimensional entity like a filament this implies that no more mass is attached 
to the filament at larger radii. In other words, the radius $r \sim 2\Mpch$ marks the 
average maximum extent of a filament. Within this range, the profile turns to a 
power-law shape with a slope $\gamma \approx -1$, which corresponds roughly to an 
isothermal profile for a filamentary entity. The profile slope transition at 
around $r \sim 2 \Mpch$, provides a simple criterion for defining the width 
of filaments. The fact that there appears to be only a small variation in this 
width (see fig.~\ref{fig:filaments_density_profile}), means that we may have some confidence 
in using this one particular value. 

To provide an impression of the variation in the density distribution around filaments, 
we show the density profile for 15 individual filaments in the right panel of the same 
figure. While the individual profiles differ, their variation is restricted to a rather 
small range. We also find some variation in the maximum extent of filaments, 
which confirms the impression obtained from the observed Cosmic Web and had already 
been noticed in previous studies \citep{Colberg05,Dolag06}. 

\begin{figure}
  \centering
  \vskip -0.5truecm
  \mbox{\hskip -1.7truecm\includegraphics[width=10.5truecm,angle=0.0]{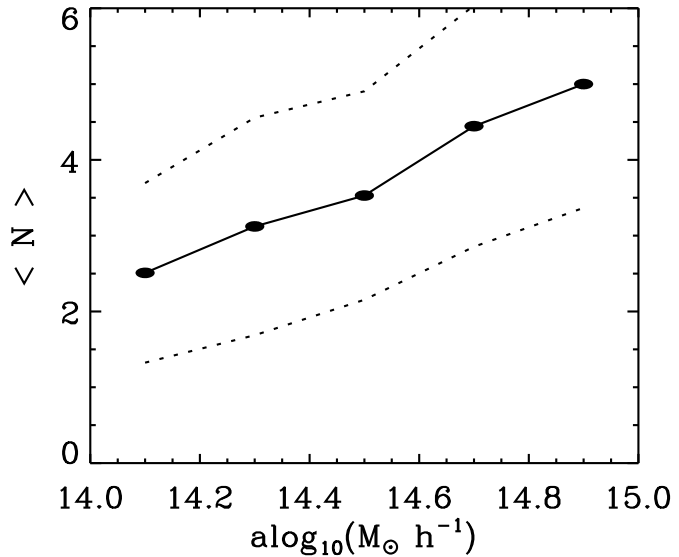}}
    \caption{Cluster Mass and Filament Connections. The figure plots the mean number of filaments, $\langle N \rangle$, 
             connected to a cluster as a function of the mass $M_{cl}$ of the cluster (solid line). The dotted line 
             indicates the $1\sigma$ dispersion of the data around the mean (see text).}
  \label{fig:cluster_N_filament_mass}
\end{figure}

\subsection{Cluster Bridges}
Filaments are closely affiliated to clusters. According to the Cosmic Web theory \citep{Bondweb96}, 
filaments are the transport channels along which clusters accrete mass \citep[see][]{Haarlem93,Diaferio97,
Colberg99}. Massive clusters are formed at the sites of rare high density peaks in the primordial matter 
distribution \citep{bbks}, and dominate the dynamical evolution of the cosmic matter distribution. Particularly 
in the high-density areas, the overwhelming coherent gravitational tidal forces between two cluster 
peaks are responsible for anisotropic collapse of the surrounding matter towards elongated filaments 
\citep{Bond96a,Bondweb96}. The strength of the filamentary bridges is expected to depend on 
the mass of the generating cluster, their mutual distance and their mutual orientation \citep{Bondweb96}. 
This has been confirmed by numerous numerical simulations \citep[see e.g.][]{Colberg05,Sousbie09,Gonzalez09}. A visual 
assessment of observations as well as simulations also suggests that the more prominent clusters are associated 
with filamentary systems of a higher complexity \citep{Pimbblet04b,Colberg99,Colberg05}. 

Figure~\ref{fig:cluster_filament_connections_particles} shows three examples of groups and 
clusters with masses in the range  $10^{13} - 10^{14}$ M$\odot$ h$^{-1}$. A first impression is that 
massive clusters appear to be embedded within a more complex filamentary environment. On the basis of 
this figure, we may make a few direct observations:
\begin{itemize}
\item The top panel shows a cluster connected to two filaments. Its mass, $M \sim 10^{13}$  M$\odot$ h$^{-1}$, 
is characteristic for a group of galaxies consisting of a few tens of galaxies. Such filaments may be the result of a  
gravitationally induced fragmentation of a longer filament. The infall pattern of matter into these clusters is highly 
anisotropic, and is mainly restricted to the direction of the connecting filaments. 

\item The cluster in the middle panel has a mass of $M \approx 2-3 \times 10^{13}$ M$\odot$ h$^{-1}$. 
It is connected to several filaments. It is also indicative that we find several other clusters in its vicinity. 

\item The cluster in the bottom panel is embedded in a highly complex environment. Several filaments
can be seen branching in a range of different directions. It is not possible to identify a main filament to
which the cluster is connected.
\end{itemize}
Following these observations along with others obtained from the simulation, we find that the number 
filamentary extensions of a cluster is closely related to the mass of the cluster.

\begin{table*}
    \begin{center}
      \begin{large}
	\begin{tabular} {l c c c c }
	  & \% Clusters & \% Filaments & \% Walls & \% Voids\\
	  \hline
	  \hline
	  Volume filling                  & 0.38 & 8.79 & 4.89 & 85.94\\
	  Mass content                    & 28.1 & 39.2 & 5.45 & 27.25\\
	  Relative density                & 73   & 4.45 & 1.11 & 0.31 \\
	  \hline
	\end{tabular}
      \end{large}
      \caption{Inventory of the Cosmic Web in terms of volume and mass content.}
      \label{tab:cosmicwebinventory}
    \end{center}
  \end{table*}
\subsubsection{Cluster Mass and Filament Connections}
In order to quantify the relation between the mass of a cluster and the number of connected  filaments 
we used the filament catalogue described in sect.~\ref{sec:length_of_filaments}. To this end, we applied 
the following criteria: 
\begin{itemize}
\item[a)] A filament is connected to a cluster if it has at least one halo within a sphere
of radius 3 $\Mpch$ from the center of the cluster. 
\item[b)] Only clusters with $M \ge 10^{14}$ M$_{\odot}$ h$^{-1}$ were considered in our analysis. 
\end{itemize}

Figure~\ref{fig:cluster_N_filament_mass} suggests an almost linear relation between 
the mass of a cluster and the number of filaments that are connected to the cluster, 
although the dispersion around the relation is rather large. The most massive clusters, the 
ones with a mass in excess of $M \sim 10^{15}$ M$_{\odot}$ h$^{-1}$ may form 
a node connecting to 5 or even 6 different filaments. Low mass clusters with a mass 
$\le 10^{14}$ M$_{\odot}$ h$^{-1}$, on the other hand, tend to have rather simple connections 
of 2 and not more than 3 filaments. 

The figure suggests that some clusters may even have only one filament connected to them. 
However, this is not observed in the Cosmic Web, and would be difficult to justify from a 
physical point of view. It is mainly the result of the method used to assign filaments to 
clusters, which may in some situations miss a few faint filaments which would be connected 
to these low-mass clusters. 

The large dispersion in figure~\ref{fig:cluster_N_filament_mass} reflects the 
complexity of the Cosmic Web. The final matter configuration in the neighborhood of 
a cluster depends not only on its mass but also on the geometrical configuration of the
surrounding clusters \citep{Bondweb96}. Other studies have found a similar relation
based on intracluster filaments found in N-body simulations \citep{Colberg05} and 
visually identified filament-cluster connections from the 2dF galaxy survey 
\citep{Pimbblet04b}.

\section{Conclusions}
\label{sec:conclusions}
We provide a qualitative and quantitative description of the Cosmic Web in terms
of its morphological constituents. We focused on filaments, to a
lesser degree on walls. The basis for this work is a large N-body
simulation of a $\Lambda$CDM cosmology with dark matter particles. 
The morphological segmentation was done with the Multiscale Morphology Filter.

\begin{enumerate}
\item[$\bullet$] The mass content, volume content and mean density of the cosmic web is 
quantitatively summarized in table~4.

\item[$\bullet$] Each morphology of the cosmic web has a characteristic density. The distribution
  of densities however, has a large overlap. Density alone can give an indication
  of the morphology but it can not be used to unambiguously segment the Cosmic Web 
  into its morphological constituents.

\item[$\bullet$] We offer a qualitative classification of filaments based on their visual properties
  in four types: \textit{line}, \textit{star}, \textit{grid} and \textit{complex}.

\item[$\bullet$] The density profile of filaments indicates that their typical radial extent is
  of the order of 2 $\Mpch$, although there are significant variations between filaments. In their 
  interior, filaments have a power-law density profile with slope $\gamma \approx -1$, corresponding 
  to an isothermal density profile. 

\item[$\bullet$] We find a relation between the mass of a cluster and the number of filaments
  it has connected. More massive clusters have more filaments in average. 
  Clusters with a mass of $\sim 10^{14}$ M$_{\odot}$ h$^{-1}$ have on average two connecting
  filaments while clusters with mass of $\sim 10^{15}$ M$_{\odot}$ h$^{-1}$ have on average five connecting
  filaments.
\end{enumerate}

Having analyzed and described the structure of the filamentary network, in the subsequent 
paper we will adress the velocity flows and the dynamics of the network and of individual 
clusters. Also, we plan to experiment with the MMF detection technique, and instead of 
adressing the multiscale character of the density field develop a version based on the 
dynamically more relevant gravitational potential field. 

In a related study \citep{Jones10}, we have applied the MMF to detecting filaments in the 
SDSS galaxy redshift survey and identified edge-on galaxies within their realm. This allowed 
us to adress and answer the question whether there are significant alignments of galaxy 
spins along the spine of the SDSS filaments. This is expected following the tidal torque 
theory for galaxy angular momentum generation. MMF indeed allowed us to identify a subset 
of galaxies and filaments conforming to a significant alignment. 

The application of MMF to real galaxy surveys introduces a few important additional 
challenges. One aspect concerns the survey volume selection. While a volume-limited survey 
would guarantee an ideal homogeneous coverage, in practice it involves a severe reduction of 
spatial resolution and hence the feasibility of identifing crucial aspects of the anisotropic 
features in the Cosmic Web. For magnitude-limited galaxy redshift surveys, the DTFE density 
field reconstruction enables a correction for the diminishing sampling density at higher density. 
Nonetheless, an appropriate MMF web analysis will still be restricted to the more densely 
sampled regions out to the peak of the radial survey selection function. A recent meticulous 
analysis by \cite{Platen10} has assured us that our web analysis tool box can be succesfully 
tuned towards controlling this issue. 

An additional artefact that may severely affect the identification of filaments is the imprint of 
redshift distortions. The large peculiar motions in and around the virialized cluster cores 
generate artificial radially directed elongated features, better known as the {\it Fingers of 
God}. In practice, we remove all filaments within a limiting angle $\eta_c$ around the 
line of sight. \cite{Jones10} show that this succesfully recovers a statistically proper filament 
distribution. Large scale cosmic flows are known to enhance the contrast of anisotropic 
filamentary features \citep[see e.g.][]{Shandarin09b}. However, a proper correction 
for this would at least demand a densely sampled environment so that one may model 
the corresponding force field. This is work in progress and has not yet been implemented 
in our tool box. 

\section*{Acknowledgements}
The authors are grateful to the (anonymous) referee for the careful and 
constructive appraisal of our manuscript, leading to substantial improvements.

\appendix
\section{Filament and Wall compression}
\label{sec:compression}
In order to enhance filaments and walls and to morph them closer to idealized 
lines and planes, we developed a \textit{compression} algorithm. The algorithm 
displaces particles in the direction of increasing density, towards the spine 
of the filaments or the planes that define the walls. The algorithm can be applied 
to particles as well as to haloes, after weighing them by their mass.

\begin{figure*}
  \centering
  \includegraphics[width=0.9\textwidth,angle=0.0]{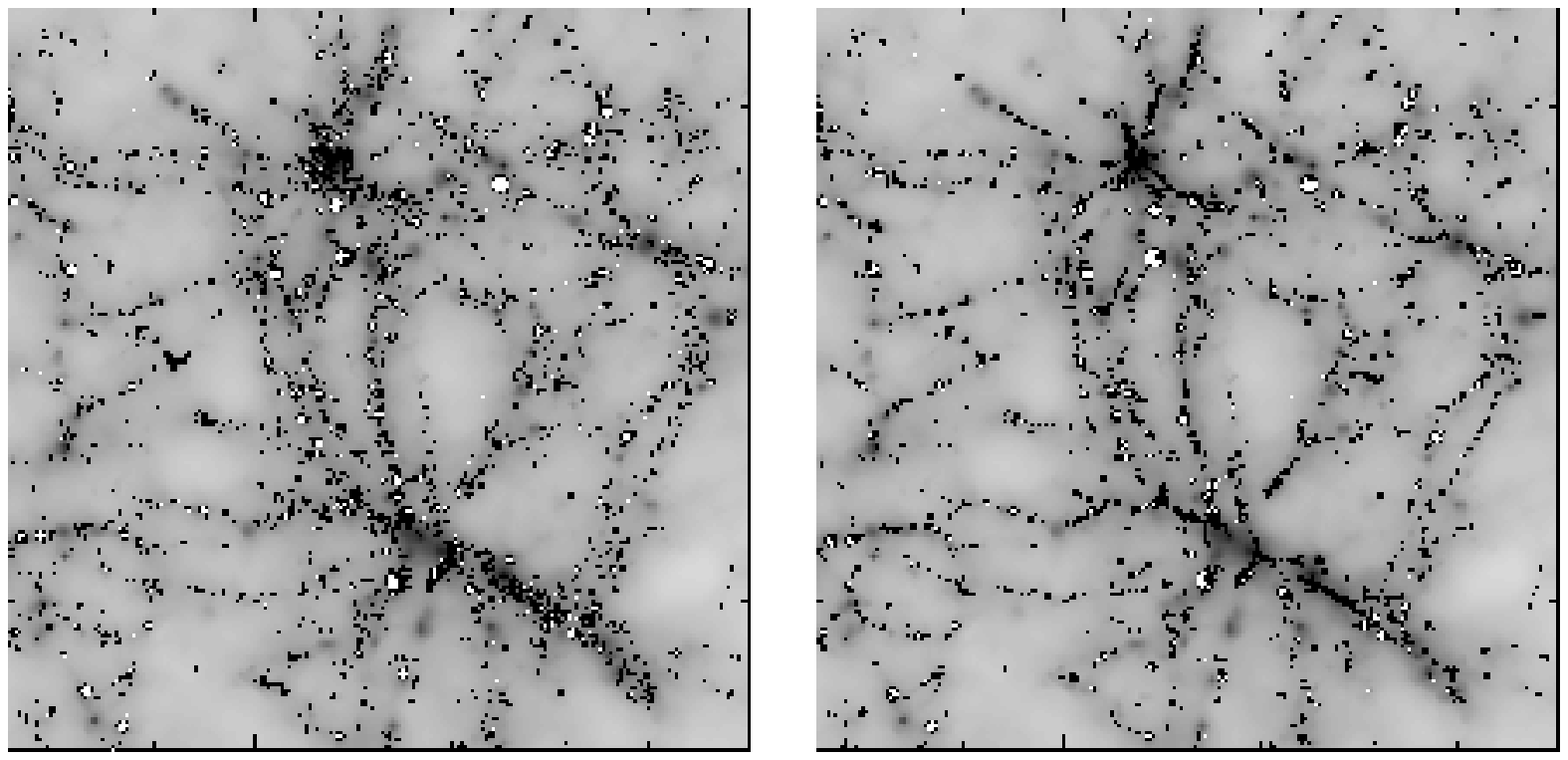}
    \caption{Filament compression illustrated. Haloes in filaments before (left) and after the application of the 
             compression algorithm (right). Haloes are indicated as white circles whose sizes are scaled according 
             to their mass. The gray background delineates the density field plotted in logarithmic scale.}
  \label{fig:slice_filas_compress_zoom}
\end{figure*}

The compression procedure exploits the information produced by the MMF. Not 
only does it identify the filaments and walls in the cosmic web, but also their   
local direction represented by the eigenvectors of the Hessian matrix 
(see \cite{Aragon07b} and sect.~\ref{sec:mmf}). The smallest eigenvector of the 
Hessian matrix traces the local direction of a filament, while the largest eigenvector 
locally defines the normal to the wall. This information is exploited to 
iteratively displace particles to the local center of mass within a given radius. 
It is a widely used method for computing the center of mass in spherical or 
semi-spherical haloes \citep{Bosch02}.

\begin{figure}
  \centering
  \includegraphics[width=0.4\textwidth,angle=0.0]{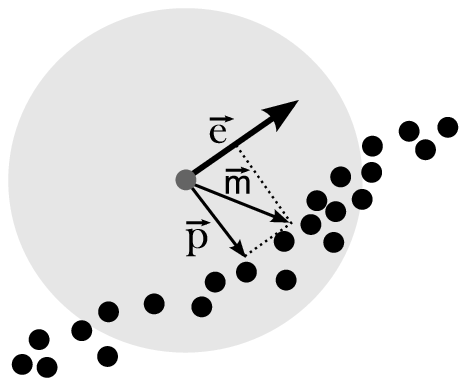}
    \caption{Cartoon illustrating the compression steps for an individual galaxy. 
             It shows how the galaxy is displaced towards a position on the 
             heartline of the filament, along a direction perpendicular to the 
             filament in which it is embedded.}
  \label{fig:compress_fil_diag}
\end{figure}

We start by defining the \textit{heartline} of filaments and walls in a similar 
way similar to determining the center of mass in spherical haloes. This heartline is used 
as reference point. The one and two-dimensional heartlines for filaments and walls are referred 
as the \textit{spine} of the filaments and the \textit{plane} of the walls.

Our compression algorithm involves the crucial constraint that the displacement of 
particles to the center of mass follows the direction \textit{perpendicular to
the filament or wall}. As a result, this process transforms thick structures into thin lines 
or planes without affecting their length. 

In summary, the algorithm has the following steps (for simplicity we only refer to 
particles, haloes will be treated equivalently): 

\begin{enumerate}
\item[$\bullet$] For each particle $i$ we find all neighbours inside a top-hat window 
of a given radius $R_{top}$, centered at the particle's position  $\bf x_i$. The
tophat radius should be large enough to enclose the filament or wall in order to minimize 
the number of iterations needed. 
\item[$\bullet$] The center of mass $\bf m_i$ of the particles inside the tophat window is computed, 
along with the vector defined by the particle's position and the center of mass 
$\bf m_i$.
\item[$\bullet$] The particle is displaced to the center of mass \textit{along the
perpendicular direction of the filament/wall}:
\begin{equation}
    \bf p = (\bf e * \bf m) * \bf e \sin(\theta)
\end{equation}
\noindent where  $\bf e$ is the vector indicating the direction of the
spine of the filament or the normal to the plane of the wall. The angle 
$\theta$ is the angle between the vector $\bf e$ and the center of mass 
$\bf m$ (see fig.~\ref{fig:compress_fil_diag}). 

The eigenvectors $\bf e$ are computed from the Hessian matrix smoothed at the 
characteristic scale of filaments, $\sim 2-3 \Mpch$ (see sect.~\ref{sec:density_profiles}).
\item[$\bullet$] After having performed the process for all particles, we 
compute the dispersion between the ``original'' and ``compressed'' positions.
We repeat the complete process until the dispersion between consecutive iterations 
changes by a given factor, or until the total dispersion is less than a prescribed 
value. This value specifies the convergence of the method.
\end{enumerate}
\noindent The compression algorithm is rather insensitive to the size of the tophat 
window, which may be in the range of $R_{th} \sim 1-5\Mpch$ for filaments and $R_{th}\sim 2-8\Mpch$ 
for walls. The lower limit depends on the mean interparticle separation, because there 
must be at least two particles inside the search window in order to displace the particle.
The maximum value of $R_{th}$ depends on how close the filaments and walls are to each 
other. If we would chose for a larger radius, particles from adjacent structures 
would be included which would invalidate the compression procedure.

We continue the compression algorithm recursively until the dispersion between 
consecutive compressed positions is less than 0.25 $\Mpch$. It is important to note 
that in the compression algorithm we only account for particles contained in the 
morphological population under consideration.\\

\section{Length of filaments}
\label{sec:fillength}
The determination of the length of filaments involves two steps. The first 
is the compression of the filaments, following the procedure outlined 
in the appendix~\ref{sec:compression}. The second step consists of fitting 
a polynomial to the particle or halo distribution along the filament.

By fitting to a polynomial, we smooth the small-scale variations that may remain 
following the compression procedure. If we would not include this step, and instead 
chose to add the segments of the mimimum spanning tree defined by the particle 
distribution, we would end up with a filament whose size would be larger than 
strictly representative. 

As a compromise between simplicity and the ability to follow intricate and 
complex filament shapes, we chose to use polynomials of $3^{rd}$ order. A 
visual inspection of several filaments assured us that the $3^{rd}$ order 
polynomials are indeed sufficient for modelling even the most intricate 
filaments. They manage to follow each change in direction. 

The positions ${\bf r}_i=(x_i,y_i,z_i)$ of each of the particles/haloes $i$ belonging to 
a filament are fitted to a polynomial of the form
\begin{eqnarray}
x = a_1 + b_1 t + c_1 t^2 + d_1 t^3\,,\nonumber\\
y = a_1 + b_1 t + c_1 t^2 + d_1 t^3\,,\\
z = a_1 + b_1 t + c_1 t^2 + d_1 t^3\,,\nonumber
\end{eqnarray}
\noindent where the parameter $t$ is defined as the distance from an arbitrary location
($x_0,y_0,z_0$):
\begin{equation}
t=\sqrt{(x-x_0)^2+(y-y_0)^2+(z-z_0)^2}
\end{equation}
\noindent In practice we chose a set of values for $(x_0,y_0,z_0)$, usually the corners of the
simulation box, and select the one that gives the best fit according to the criterion of 
having the smallest mean square difference,
\begin{equation}
   \epsilon = \frac{1}{N} \chi^2.
\end{equation}
\noindent If $\epsilon$ turns out to be larger than a given threshold, whose value is 
determined by means of experimentation, we reject the fit. 

\begin{figure}
  \centering
  \includegraphics[width=0.45\textwidth,angle=0.0]{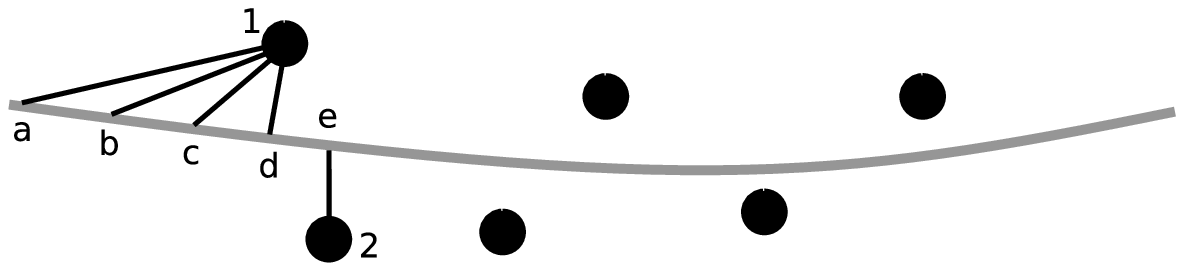}
  \caption{Identification of extreme of a filament: The fitting curve (gray line) is closely
  sampled at points $a,b,c,d,e$. Haloes are represented as large black dots. The closest halo 
  to points $a,b,c$ and $d$ is halo 1. At point $e$ the closest halo changes to halo 2 indicating 
  that the fitting curve is ``inside'' the filament.}
  \label{fig:filaments_compressed_fit}
\end{figure}

One remaining technical difficulty is the determination of the extreme points of the filament 
for the fitting curve. This is a non-trivial task, and if not considered properly may lead to
wrong length determinations. We follow a simple but effective method to identify the
extremes:
\begin{itemize}
\item[$\bullet$] The polynomial curve is closely sampled and distances to all the particles in the
  filament are computed starting from one extreme of the fitted curve.
\item[$\bullet$] For each point in the fitting curve the closest halo is identified.
\item[$\bullet$] We identify the point in the fitting curve at which the identity of the closest
halo changes. This indicates that the fitting curve is no longer ``outside'' the
filament but ``inside'' it (see fig.~\ref{fig:filaments_compressed_fit}).
\item[$\bullet$] We repeat the procedure for the second extreme of the fitting curve.
\end{itemize}

\noindent Following the previous steps, the fitted polynomial is used to compute the length of the filament. 
\end{document}